\documentclass{article} 
\usepackage[preprint]{neurips_2025}
\makeatletter
\renewcommand{\@noticestring}{Preprint.}
\makeatother

\usepackage{amsmath,amsfonts,bm}

\def\eqref#1{equation~\ref{#1}}

\def\1{\bm{1}}

\DeclareMathAlphabet{\mathsfit}{\encodingdefault}{\sfdefault}{m}{sl}
\SetMathAlphabet{\mathsfit}{bold}{\encodingdefault}{\sfdefault}{bx}{n}

\usepackage[colorlinks=true,linkcolor=red!60!black,citecolor=blue!60!black,urlcolor=magenta!60!black]{hyperref}
\usepackage{url}
\usepackage{booktabs}
\usepackage{xcolor}
\usepackage{multirow}
\usepackage{amsthm}
\usepackage[utf8]{inputenc}
\usepackage[T1]{fontenc}
\usepackage{textcomp}
\usepackage{graphicx}
\usepackage{xcolor}
\usepackage{listings}
\usepackage[most]{tcolorbox}
\usepackage{subcaption}
\usepackage{amssymb}   
\DeclareUnicodeCharacter{2212}{\textminus}
\newcommand{\std}[1]{\,\textcolor{gray}{\scriptsize$\pm$#1}}
\newcommand{\best}[1]{\textbf{#1}}
\newcommand{\second}[1]{\underline{#1}}
\newcommand{\flowreader}{\textsc{FlowReader}}
\newtheorem{proposition}{Proposition}

\title{Min-Cost Flow Routing for Evidence Assembly in Long Multimodal Documents}

\author{%
  Ambuj Mehrish \\
  Ca' Foscari University of Venice \\
  \texttt{ambuj.mehrish@unive.it} \\
  \And
  Sebastiano Vascon \\
  Ca' Foscari University of Venice \\
  \texttt{sebastiano.vascon@unive.it} \\
}

\begin{document}

\maketitle

\begin{abstract}
Answering questions about long multimodal documents requires distributing a fixed evidence budget across relevant facets in text, tables, figures, and slides while avoiding near-duplicates. We present \flowreader, which formulates evidence selection as a single minimum-cost flow problem with capacity limits over a multimodal content graph. Spectral decomposition identifies latent aspects of query-relevant content and allocates the budget among them in proportion to their spectral energy. These capacity limits enforce aspect coverage during routing without requiring a language-model planning call. Query-conditioned costs prioritize chains of relevant, mutually consistent evidence. Decomposing the optimal flow produces short evidence chains, which a vision-language model reads in parallel and a reasoner reconciles. On VisDoMBench with Qwen3-VL-32B, \flowreader\ achieves the highest macro accuracy ($68.9$), surpassing the strongest prior system by $2.7$ points, leading on three of five subsets and attaining the highest worst-subset accuracy. It uses a measured $17.5$ content nodes per query and maintains its lead at $12.9$. Ablation studies with a fixed graph, scorer, reader, and judge show that cost design drives accuracy, capacity limits preserve it while using about three-quarters of the reader tokens required by shortest-path routing without these limits on the same network, and spectral aspects align with LLM-generated sub-questions without a planning call\footnote{Code will be released soon.}.

\end{abstract}

\section{Introduction}
\label{sec:intro}
Multimodal documents distribute information across text, tables, figures, captions, and pages, often requiring readers to combine content with its surrounding context. Retrieval-augmented generation (RAG) supplies evidence beyond a language model's immediate context~\citep{lewis2020retrieval}. However, conventional text retrieval often overlooks document structure: fixed-length chunks can separate relevant content from the context needed to interpret it~\citep{wang2025document,suri2025visdom,dong2024mc,ma2024mmlongbench,gao2026scaling}. Ranking retrieval units independently compounds this problem because a unit's value can depend on what is retrieved alongside it~\citep{li2024graphreader,gutierrez2024hipporag}. A table header, numerical cell, or earlier slide may score poorly in isolation yet be essential for interpreting another unit. We refer to this separation of interdependent content as \emph{fragmented evidence}.

Selecting fragmented evidence under a fixed budget requires coordination. Choosing the highest-scoring units can favour the query's most similar facet, retrieving near-duplicate evidence while omitting lower-scoring units needed for a complete answer. Redundant evidence also consumes the reader's limited context~\citep{liu2024lost}. Existing methods address different aspects of this problem: multimodal retrievers operate on page images or regions~\citep{faysse2024colpali,yuvisrag,suri2025visdom,wang2025vidorag,xu2025multigranularity}; graph-based RAG methods connect evidence units~\citep{edge2024local,guo2024lightrag,wan2025mmgraphrag,yan2026question,du2026g}; and joint selection methods use submodular, diversity, or transport objectives~\citep{lin2011class,carbonell1998use,kulesza2012determinantal,cuturi2013sinkhorn}. Recent work also considers coverage and retrieval cost in RAG~\citep{zhang2026georag,wu2026cost}.

G\textsuperscript{2}-Reader uses a large language model (LLM) planner to generate sub-questions, retrieve evidence for each, and assess whether the evidence is sufficient~\citep{du2026g}, following an iterative retrieval approach~\citep{trivedi2023interleaving,asai2023self}. Such planning makes evidence needs explicit but does not guarantee an optimal allocation of the retrieval budget: evidence retrieved for different sub-questions may overlap, while sequential planning adds latency.

We introduce \flowreader, which represents query decomposition through capacity constraints in a flow network. Spectral decomposition identifies latent aspects of query-relevant document content, each represented by a node with capacity proportional to its spectral energy. A single minimum-cost flow routes a fixed demand across these aspects. Node scores determine entry and termination costs, while edge costs favour transitions to relevant evidence that agrees with preceding nodes. Since the total demand equals the sum of aspect capacities, every feasible solution must satisfy the prescribed allocation. Coverage is therefore a feasibility condition rather than a soft preference (Proposition~\ref{prop:coverage}). The optimal flow is decomposed into concise evidence chains, which a vision-language model reads in parallel before a reasoner combines the findings into an answer. Unlike flow-based graph RAG methods that prune relational paths~\citep{chen2025pathrag} or diffuse query mass~\citep{zhou2026qafd}, \flowreader\ explicitly allocates retrieval demand across latent query aspects without an LLM planning call.

On VisDoMBench, \flowreader\ achieves a macro accuracy of $68.9$, exceeding the strongest prior system by $2.7$ points. It leads on three of five subsets and achieves the highest worst-subset accuracy. The system reads an average of $17.5$ distinct content nodes per query and retains a $1.8$-point lead when this is reduced to $12.9$ nodes. Its retrieval and answering pipeline uses only open-source models, including an open embedder in place of G\textsuperscript{2}-Reader's proprietary component. With the graph, node scores, reader, and judge held fixed, ablations show that node credit and the agreement objective improve accuracy by $2.3$ and $3.4$ points, respectively ($p<0.01$). Capacity constraints reduce reading cost while preserving accuracy: shortest-path routing on the same network without these constraints achieves comparable accuracy but produces chains twice as long. \flowreader\ uses $73\%$ of the reader tokens required by this alternative. Spectral aspects achieve accuracy within $0.45$ points of LLM-generated sub-questions ($p=0.35$), without a planning call.

Our contributions are: \textbf{(1)} an exact minimum-cost flow formulation for multimodal evidence selection, with aspect capacity constraints that enforce the prescribed coverage in every feasible solution; \textbf{(2)} a spectral construction of aspects and capacities that achieves comparable accuracy to LLM-generated sub-questions without a decomposition call; and \textbf{(3)} state-of-the-art accuracy on VisDoMBench, supported by controlled ablations showing that cost design improves accuracy and capacity constraints preserve it while reducing reader token use to $73\%$ of unconstrained routing. We report costs in content nodes, tokens, and model calls.

\section{Related Work}
\label{sec:rw}

\paragraph{Multimodal and graph-structured RAG.}
Recent systems retrieve evidence directly from page images or multimodal regions. ColPali~\citep{faysse2024colpali} uses late interaction over visual tokens, while VisRAG~\citep{yuvisrag}, M3DocVQA~\citep{cho2025m3docvqa}, VisDoMRAG~\citep{suri2025visdom}, and ViDoRAG~\citep{wang2025vidorag} extend visual retrieval to multi-page and multi-document question answering. RAG-Anything~\citep{guo2025rag} integrates text, tables, and images using layout-aware parsing~\citep{wang2024mineru}. Relevant benchmarks include VisDoMBench~\citep{suri2025visdom}, SlideVQA~\citep{tanaka2023slidevqa}, SPIQA~\citep{pramanick2024spiqa}, and UDA~\citep{hui2024uda}, which incorporates FeTaQA~\citep{nan2022fetaqa} and QASPER~\citep{dasigi2021dataset}.

Graph-based RAG introduces explicit relationships between evidence units. GraphRAG~\citep{edge2024local} and LightRAG~\citep{guo2024lightrag} construct entity--relation graphs, HippoRAG~\citep{gutierrez2024hipporag} applies personalised PageRank to a knowledge graph, and MMGraphRAG~\citep{wan2025mmgraphrag} incorporates multimodal nodes. G\textsuperscript{2}-Reader~\citep{du2026g} maintains two evolving graphs consolidated through VLM message passing, an approach related to graph memories for long-horizon agents~\citep{chhikara2025mem0,xu2025mem,rasmussen2025zep}. Flow-based methods use graph structure to prune relational paths~\citep{chen2025pathrag} or diffuse query mass with retrieval guarantees~\citep{zhou2026qafd}.

\paragraph{Adaptive control and optimization-based selection.}
Iterative retrieval alternates evidence acquisition with reasoning~\citep{press2023measuring,yao2022react,trivedi2023interleaving,jiang2023active,shao2023enhancing,asai2023self}. Adaptive methods adjust retrieval depth or query decomposition to query complexity~\citep{jeong2024adaptive,yu2024auto,nguyen2025ma}, while dual-process designs combine fast and reflective stages~\citep{wang2024speculative,cheng2025dualrag,hu2026cog}. \citet{liang2025reasoning} survey these approaches. Such methods use model calls to guide retrieval decisions; \flowreader\ determines the allocation in a single optimization before reading any evidence.

Evidence selection has also been formulated using integer linear programming~\citep{mcdonald2007study,gillick2009scalable}, submodular maximization~\citep{lin2010multi,lin2011class}, MMR~\citep{carbonell1998use}, and determinantal point processes~\citep{kulesza2012determinantal,cho2019improving}. RAG-specific approaches include SMART-RAG~\citep{li2024smart} and demand-coverage objectives~\citep{zhang2026georag}. Other formulations use optimal transport~\citep{kusner2015word,cuturi2013sinkhorn,tang2022otextsum,shaaban2025motor} or stochastic passage selection~\citep{zamani2024stochastic,meng2025ranking}. Replicator dynamics provide a continuous relaxation of clique selection~\citep{pelillo1998replicator,bomze2000approximating}, motivating our quality--diversity path selector.

Many selection objectives reward coverage or penalise redundancy through terms that compete with other scores. \flowreader\ instead encodes a prescribed allocation across spectral aspects through network capacities and computes an exact minimum-cost flow. When the full demand is routed, every feasible solution satisfies this allocation. This guarantee applies to the complete routed flow over spectral aspects; it does not guarantee semantic coverage or preservation of the allocation after path selection.
\section{Methodology}
\label{sec:methodology}

\flowreader\ assembles evidence through a single minimum-cost flow optimization over a query-conditioned document graph. Given a query $q$, arc capacities allocate a fixed retrieval demand across latent aspects of relevant content. The optimal flow selects and orders content nodes into evidence chains, which are read in parallel and combined into an answer. Routing and selection are completed before reading, with no replanning or dependencies between reading calls. Implementation details appear in Appendix~\ref{app:implementation}.

\begin{figure}[t]
    \centering
    \includegraphics[width=\textwidth]{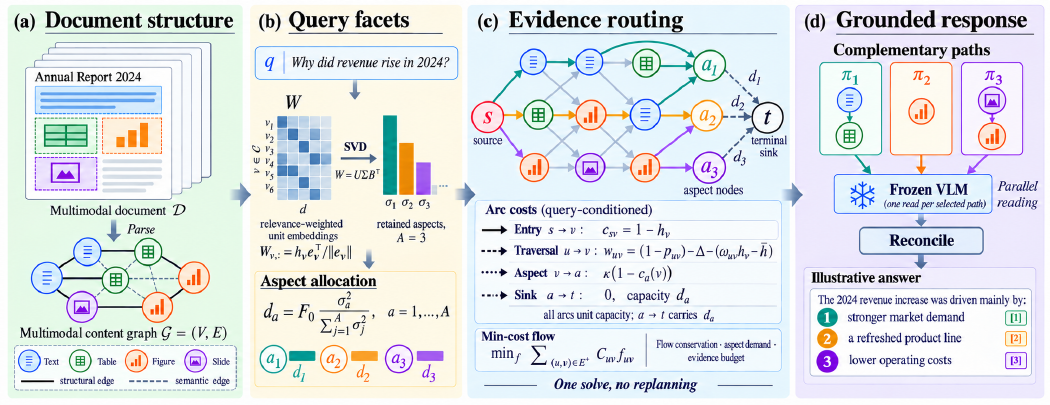}
    \caption{\textbf{Overview of \flowreader.}
    \emph{(a)} Each document is parsed once into a multimodal content graph.
    \emph{(b)} SVD of relevance-weighted node embeddings identifies latent aspects and divides the demand $F_0$ into aspect capacities $d_a$.
    \emph{(c)} A single minimum-cost flow routes this demand from source $s$ through content and aspect nodes to sink $t$. Aspect capacities enforce the prescribed allocation without a planning call.
    \emph{(d)} The flow is decomposed into evidence chains, which a VLM reads in parallel before a reasoner combines the findings.}
    \label{fig:overview}
\end{figure}

\subsection{Multimodal Graph and Query Relevance}
\label{sec:method-graph}

We represent the document collection as a graph $G=(V,E)$ whose nodes correspond to passages, tables, figures, and slides. Each node stores its content, summary, modality, and embedding $\mathbf e_v$. We adopt the offline graph construction of G\textsuperscript{2}-Reader~\citep{du2026g}, retaining its node-analysis and linking prompts except for the visual-node prompt, which we modify to include in-image text in the node's keywords. The graph is constructed once per collection and reused across queries. During routing, we add symmetric nearest-neighbor edges; edge types and neighborhood size are specified in Appendix~\ref{app:impl-graph}.

For query embedding $\mathbf q$, we define the label-aware relevance score and its max-normalized form as
\begin{equation}
 r_v=\operatorname{clip}\!\bigl(
       \cos(\mathbf q,\mathbf e_v)
       +\rho_{\mathrm{ref}}\,\ell_v(q),\;0,\;1\bigr),
 \qquad
 h_v=\frac{r_v}{\max_{u\in V}r_u},
 \label{eq:fr-relevance}
\end{equation}
where $\ell_v(q)\in[0,1]$ measures how closely node $v$ matches a figure or table label explicitly named in the query. Partial matches receive proportionally smaller boosts, and nodes without a matching reference receive none. The parameter $\rho_{\mathrm{ref}}$ controls the boost strength. We use $h_v$ throughout routing to determine entry costs, contribute to traversal costs, and construct the spectral aspects and their capacities.

\subsection{Query-Conditioned Costs and Aspects}
\label{sec:method-costs}

\paragraph{Traversal costs.}
A useful transition leads to relevant evidence that adds information beyond the preceding node. We assign each edge a utility $p_{uv}$ that combines alignment with the residual query, shared rare terms, and a bonus for transitions between visual and non-visual content (Appendix~\ref{app:impl-edges}). Its traversal cost is
\begin{equation}
 w_{uv}=(1-p_{uv})-\Delta
         -\bigl(\omega_{uv}h_v-\overline h\bigr),
 \label{eq:fr-arc-cost}
\end{equation}
where $\overline h=|V|^{-1}\sum_{v\in V}h_v$ is the mean node score, $\Delta$ is a constant offset that sets the cheapest hop cost to zero before node credit, and $\omega_{uv}$ measures embedding agreement between the endpoints. The final term provides a \emph{node credit}: it lowers the cost when agreement-weighted destination relevance exceeds the mean node score. A chain of relevant, consistent evidence can therefore cost less than a single passage.

We orient content edges toward strictly higher relevance, so chains progress from supporting context to more relevant nodes. This orientation makes the network acyclic and prevents negative-cost cycles despite the node credit. Most chains contain a single node ($75.8\%$; mean length $1.46$; Appendix~\ref{app:chains}).

\paragraph{Nominal flow demand.}
We derive the nominal demand from the participation ratio of thresholded relevance scores:
\begin{equation}
 z_v=\max(0,h_v-\tau),
 \qquad
 F_0=\operatorname{clip}\!\left(
 \frac{\left(\sum_{v\in V}z_v\right)^2}
      {\sum_{v\in V}z_v^2},
 F_{\min},F_{\max}
 \right).
 \label{eq:fr-demand}
\end{equation}
The threshold $\tau$ suppresses low-relevance nodes, while $F_{\min}$ and $F_{\max}$ bound the demand. Because $h_v$ is max-normalized, $\tau$ is relative to the highest node score. At the reported operating point, the upper bound is active for $99.7\%$ of queries, making $F_0$ effectively a fixed budget comparable to that of a fixed-$k$ retriever. However, demand measures units of flow rather than distinct content nodes, since a chain may contain several nodes.

\paragraph{Spectral aspects.}
We derive aspects from query-relevant document content rather than generated sub-questions. Let $\mathcal C=\{v\in V:h_v\ge\tau\}$ be the relevant node set. We form a matrix $W\in\mathbb R^{|\mathcal C|\times d}$ whose rows are relevance-weighted unit embeddings,
$W_{v,:}=h_v\mathbf e_v^{\top}/\lVert\mathbf e_v\rVert$.
Its singular value decomposition is $W=U\Sigma B^{\top}$, with singular values $\sigma_1\ge\sigma_2\ge\cdots$ and right singular vectors $\mathbf b_a$. These vectors define directions of variation in the relevance-weighted content, with energy given by the squared singular values. We orient each direction toward the normalized query $\hat{\mathbf q}=\mathbf q/\lVert\mathbf q\rVert$:
\begin{equation}
 \boldsymbol\phi_a=
 \begin{cases}
  \mathbf b_a,
    & \langle\mathbf b_a,\hat{\mathbf q}\rangle\ge0,\\
  -\mathbf b_a,
    & \text{otherwise}.
 \end{cases}
 \label{eq:fr-aspect-dir}
\end{equation}
Node $v$ aligns with aspect $a$ according to
$c_a(v)=\max\bigl(0,\cos(\mathbf e_v,\boldsymbol\phi_a)\bigr)$.
We retain enough aspects to meet an energy threshold $\eta$, subject to a demand-dependent cap:
\begin{equation}
 A_{\eta}=\min\left\{k:
 \frac{\sum_{j=1}^{k}\sigma_j^2}
      {\sum_j\sigma_j^2}\ge\eta
 \right\},
 \qquad
 A=\min\bigl(A_{\eta},\operatorname{round}(F_0),|\mathcal C|\bigr).
 \label{eq:fr-aspect-count}
\end{equation}
The retained spectral energies determine the aspect capacities:
\begin{equation}
 d_a=F_0\frac{\sigma_a^2}{\sum_{j=1}^{A}\sigma_j^2},
 \qquad
 \sum_{a=1}^{A}d_a=F_0.
 \label{eq:fr-aspect-capacity}
\end{equation}
This construction defines both the aspect directions and their demand allocation without an LLM planning call.

\subsection{Minimum-Cost Evidence Assembly}
\label{sec:method-routing}

We augment the graph with a source $s$, one aspect node $a$ per retained direction, and a sink $t$. Every content node can start or end a path. A source-to-content arc $(s,v)$ has cost $1-h_v$, a content arc $(u,v)$ has cost $w_{uv}$, and a content-to-aspect arc $(v,a)$ has cost $\kappa(1-c_a(v))$, where $\kappa$ controls the penalty for terminal misalignment. These arcs all have unit capacity. Each aspect connects to $t$ through a zero-cost arc with capacity $d_a$.

Let $G^+=(V^+,E^+)$ denote this network, with arc costs $C_{uv}$ and capacities $U_{uv}$. For a feasible routed demand $F\le F_0$, we solve
\begin{equation}
 \begin{aligned}
 f^{\star}\in\operatorname*{arg\,min}_{f}\quad
     &\sum_{(u,v)\in E^+}C_{uv}f_{uv}\\
 \text{subject to}\quad
     &\sum_{v:(u,v)\in E^+}f_{uv}
      -\sum_{v:(v,u)\in E^+}f_{vu}=b_u,
      &&u\in V^+,\\
     &0\le f_{uv}\le U_{uv},
      &&(u,v)\in E^+,
 \end{aligned}
 \label{eq:fr-flow}
\end{equation}
where $b_s=F$, $b_t=-F$, and $b_u=0$ otherwise. If the network cannot carry $F_0$, we reduce the demand to $F=\min(F_0,0.99\,f_{\max})$, where $f_{\max}$ is the network's maximum flow. This fallback is never required on the evaluation queries. The optimization jointly determines path origins, intermediate evidence, and terminal aspects.

When $F=F_0$, the aspect capacities sum to the routed demand, forcing every feasible solution to saturate every aspect-to-sink arc. The formulation therefore enforces the prescribed allocation:

\begin{proposition}[Exactness and coverage]
\label{prop:coverage}
(i) Problem~\ref{eq:fr-flow} is a linear program over a network polytope and is solvable exactly in polynomial time. Its constraint matrix is totally unimodular, so an integral optimum exists when the capacities and node balances are integral.
(ii) $G^+$ is acyclic, so any feasible flow decomposes into at most $|E^+|$ source-to-sink paths, with no cycles.
(iii) In any feasible flow with $F=F_0$, each aspect $a$ receives exactly $d_a$ units through at least $\lceil d_a\rceil$ distinct content nodes.
\end{proposition}

These properties follow from standard network-flow results~\citep{ahuja1993network}; proofs appear in Appendix~\ref{app:proofs}. The coverage guarantee concerns spectral directions in query-relevant content, rather than the semantic facets required to answer the question. Capacities determine \emph{how much} flow each aspect receives, while the terminal cost $\kappa(1-c_a(v))$ guides \emph{which} nodes supply it. Neither mechanism prevents semantically similar nodes from serving the same aspect. Separately, the node credit rewards relevance and agreement along a chain. When $F<F_0$, aspect capacities still limit concentration, but need not all be saturated.

We decompose the optimal flow into source-to-sink paths and remove the auxiliary nodes to obtain candidate evidence chains $\Pi=\{\pi_1,\ldots,\pi_K\}$, each with an associated flow amount.

\subsection{Path Selection and Answer Generation}
\label{sec:method-selection}

Candidate paths may share content nodes. We therefore select a compact subset for reading using either routed flow or a quality--diversity criterion. The aspect constraints apply to the complete routed flow; the selected subset is not guaranteed to preserve that allocation.

\paragraph{Greedy selection.}
We rank candidate paths by their routed flow and retain the top $B\le K$, where $B$ is the path-selection budget. This rule uses the optimization's flow amounts directly, without an additional path-scoring model.

\paragraph{Replicator dynamics.}
Alternatively, we select chains using annealed replicator dynamics with a payoff that rewards relevance and low node overlap (Appendix~\ref{app:rd}). The reported system uses this approach. Greedy selection is sufficient when routed flow already ranks candidates well, while replicator dynamics provides the largest gains when the candidate pool is large and contains redundant chains (Section~\ref{sec:ablate}).

\paragraph{Answer generation.}
We follow the worker--reasoner design of G\textsuperscript{2}-Reader~\citep{du2026g}, with evidence units formed by flow routing. A vision-language worker reads each selected chain's text and images and produces a draft answer to the original query, together with supporting evidence. G\textsuperscript{2}-Reader's workers instead answer planner-generated sub-questions in dependency order, followed by a sufficiency check. In \flowreader, all chains are read in a single parallel round.

A reasoner receives the drafts and their evidence, combines complementary findings, and resolves conflicts against the source content (Appendix~\ref{app:prompts}). An additional call identifies the question type and expected answer length to set the reasoner's evidence limits; it does not decompose the query. Routing and path selection depend only on the query and document graph, never on the results of earlier reading calls.
\section{Experiments}
\label{sec:results}


\begin{table*}[t]
\centering
\small
\setlength{\tabcolsep}{5.5pt}
\renewcommand{\arraystretch}{1.1}
\resizebox{0.8\columnwidth}{!}{%
\begin{tabular}{@{}l ccccc @{\hspace{10pt}} c@{}}
\toprule
\textbf{Method} & \textbf{SPIQA} & \textbf{FetaTab} & \textbf{PaperTab} &
\textbf{SciGraphQA} & \textbf{SlideVQA} & \textbf{Average} \\
\midrule
\addlinespace[1pt]
\multicolumn{7}{@{}l}{\textit{Single-VLM, no retrieval}} \\
\addlinespace[1pt]
GPT-5
  & 55.2\std{0.1} & 63.9\std{0.3} & 37.1\std{0.1} & 64.1\std{0.3} & 45.1\std{0.1} & 53.1\std{0.1} \\
Qwen3-VL-32B
  & 29.9\std{0.1} & 37.4\std{0.4} & 34.3\std{0.3} & 23.1\std{0.2} & 24.9\std{0.2} & 29.9\std{0.1} \\
\addlinespace[3pt]
\midrule
\addlinespace[3pt]
\addlinespace[1pt]
\multicolumn{7}{@{}l}{\textit{Retrieval-augmented generation} (Qwen3-VL-32B-Instruct)} \\
\addlinespace[1pt]
MA-RAG
  & 45.5\std{0.2} & 27.7\std{0.2} & 33.4\std{0.5} & 29.3\std{0.3} & 29.4\std{0.2} & 33.1\std{0.1} \\
GraphRAG
  & 62.7\std{0.2} & 61.4\std{0.2} & 42.9\std{0.0} & \second{65.8}\std{0.4} & 21.7\std{0.0} & 50.9\std{0.1} \\
RAGAnything
  & 67.7\std{1.0} & 57.8\std{0.2} & 42.0\std{1.4} & 41.6\std{2.6} & 52.2\std{0.5} & 52.3\std{0.6} \\
ViDoRAG
  & 68.2\std{0.5} & 58.7\std{0.4} & 43.7\std{0.2} & 37.9\std{0.1} & \second{71.7}\std{0.1} & 56.0\std{0.1} \\
Deepseek-OCR
  & 63.6\std{0.4} & 70.3\std{0.1} & 51.6\std{0.2} & 61.9\std{0.4} & 65.7\std{0.1} & 62.6\std{0.1} \\
MMGraphRAG
  & 69.9\std{0.2} & \second{72.4}\std{0.6} & 56.4\std{0.6} & 64.1\std{0.3} & 54.2\std{0.2} & 63.4\std{0.2} \\
LightRAG
  & 73.9\std{0.0} & 64.7\std{0.4} & 51.0\std{0.0} & \best{75.0}\std{0.0} & 29.6\std{0.0} & 63.7\std{0.1} \\
VisDoMRAG
  & \second{75.4}\std{0.0} & 61.0\std{0.5} & 56.2\std{0.2} & 63.4\std{0.1} & 69.0\std{0.4} & 65.0\std{0.1} \\
G\textsuperscript{2}-Reader
  & 73.2\std{0.2} & 66.9\std{0.1} & \second{57.1}\std{0.2} & 61.6\std{0.1} & \best{72.3}\std{0.0} & \second{66.2}\std{0.1} \\
\addlinespace[3pt]
\midrule
\addlinespace[1pt]
\textbf{FlowReader} (ours)
  & \best{77.7}\std{0.3} & \best{74.0}\std{0.8} & \best{62.3}\std{0.2} & 61.8\std{0.4} & 68.8\std{0.9} & \best{68.9}\std{0.4} \\
\bottomrule
\end{tabular}}
\caption{\textbf{Accuracy (\%) on VisDoMBench.} Baseline results are transcribed from \citet{du2026g} and utilize the Qwen3-VL-32B-Instruct backbone, judge, and evaluation prompt. The reported average for LightRAG corresponds to the value in \citet{du2026g} (mean of subset scores: $58.8$). We report mean $\pm$ standard deviation over three independent runs. \best{Bold}: best; \second{underline}: second best.}
\label{tab:main}
\end{table*}

\subsection{Setup}
\label{sec:setup}

\paragraph{Dataset and baselines.}
We follow the VisDoMBench~\citep{suri2025visdom} evaluation protocol and use the data released in the public repository of \citet{du2026g}.\footnote{\url{https://github.com/DorothyDUUU/G2_Reader/tree/main}}
The benchmark comprises five subsets: FetaTab (Wikipedia tables), PaperTab (scientific tables), SPIQA (scientific figures, tables, and text), SciGraphQA (academic graphs), and SlideVQA (slide decks). Each question is paired with five documents containing relevant evidence and distractors. We evaluate all released queries, scoring duplicate questions in the SciGraphQA file once; this leaves the mean unchanged. Baseline results are taken from \citet{du2026g}.

\paragraph{Implementation and evaluation.}
\flowreader\ uses MinerU~\citep{wang2024mineru} for document parsing, $768$-dimensional nomic-embed-text embeddings~\citep{nussbaum2024nomic}, and Qwen3-VL-32B-Instruct~\citep{bai2025qwen3}, served through vLLM at temperature $0$. G\textsuperscript{2}-Reader instead uses $1{,}536$-dimensional text-embedding-3-small embeddings. We score answers using GPT-4o-mini at temperature $0$ with the exact evaluation prompt of \citet{du2026g}, given in \hyperref[app:prompt:c1]{Prompt P7}, ensuring that our results and the reported baselines use the same rubric. We report means and standard deviations over three independent runs launched on different days. Judge agreement across repeated scoring is $96$--$99\%$. Further protocol details appear in Appendix~\ref{app:protocol}.

\paragraph{Retrieval budget.}
Following \citet{du2026g}, we express retrieval cost in content nodes and additionally report reader tokens and model calls. \flowreader\ reads $16$ chains per query, covering an average of $17.5$ distinct nodes and requiring $18$ VLM calls. G\textsuperscript{2}-Reader retrieves up to $k=5$ nodes per sub-question, with up to six sub-questions, or eight after refinement~\citep{du2026g}. These settings give a configured maximum of $30$ or $40$ nodes, respectively.

\paragraph{Hyperparameters.}
Following \citet{du2026g}, we conduct ablations on the same $250$ randomly sampled examples released with G\textsuperscript{2}-Reader, drawn from the five subsets. We set the demand cap to $F_{\max}=16$ before any sweep, confirmed this choice on these examples, and held it fixed across all datasets and variants except the demand sensitivity analysis. Table~\ref{tab:demand} reports this analysis on the full evaluation set: reducing the budget by $26\%$ lowers macro accuracy by $0.9$ points. We sweep each of the three remaining constants, $\eta$, $\rho_{\mathrm{ref}}$, and $b$, on the same $250$ examples (Table~\ref{tab:constants}).

\subsection{Main results}
\label{sec:mainres}

\paragraph{Overall performance.}
Table~\ref{tab:main} reports accuracy on VisDoMBench. \flowreader\ achieves a macro average of $68.9\pm0.4$, outperforming the strongest prior system, G\textsuperscript{2}-Reader, by $2.7$ points. It leads on SPIQA ($77.7$), FetaTab ($74.0$), and PaperTab ($62.3$). The pipeline uses only open-source models, including $768$-dimensional nomic-embed-text embeddings, whereas G\textsuperscript{2}-Reader uses proprietary $1{,}536$-dimensional OpenAI text-embedding-3-small embeddings. Only the evaluation judge, shared across all baselines, is proprietary. Prior systems show uneven performance across subsets. LightRAG leads on SciGraphQA but scores $29.6$ on SlideVQA, while G\textsuperscript{2}-Reader leads on SlideVQA but scores $57.1$ on PaperTab. \flowreader\ achieves the highest worst-subset accuracy ($61.8$), indicating stronger performance on its weakest document type. It also exceeds the same backbone answering directly from pages by $39.0$ macro accuracy points.

\paragraph{Subset results and computational cost.}
\flowreader\ exceeds the best prior score on PaperTab by $5.2$ points, SPIQA by $2.3$, and FetaTab by $1.6$. These subsets often require combining tables or figures with supporting text. The macro standard deviation across runs is $0.4$ points, small relative to the $2.7$-point average gain. Each query requires $18$ VLM calls: $16$ parallel chain reads, one reasoner call, and one question-type call. A single SVD and an exact minimum-cost flow allocate retrieval demand across latent query aspects, without planning, refinement, or sufficiency checks.

\begin{figure}[t]
  \centering
  \includegraphics[width=0.5\linewidth]{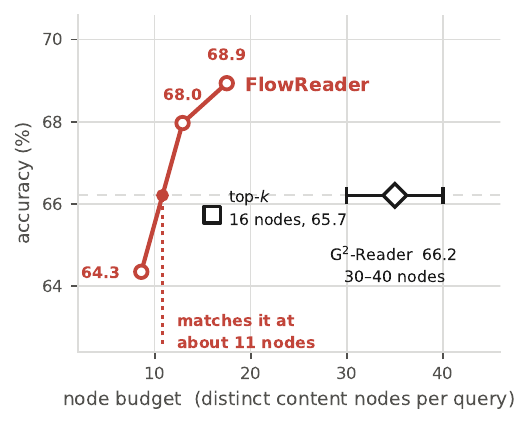}
  \caption{\textbf{Accuracy against the measured node budget.} \flowreader\ at three demands meets G\textsuperscript{2}-Reader's reported accuracy (dashed) at about eleven nodes. G\textsuperscript{2}-Reader's marker spans its configured budget of $30$--$40$ nodes~\citep{du2026g}.}
  \label{fig:budget}
\end{figure}

\paragraph{Performance at smaller budgets.}
Figure~\ref{fig:budget} plots accuracy against the measured node budget as flow demand decreases. At $12.9$ nodes, \flowreader\ achieves $68.0$ accuracy, retaining a $1.8$-point lead over G\textsuperscript{2}-Reader. Interpolation between measured budgets suggests that it matches G\textsuperscript{2}-Reader's reported accuracy at approximately eleven nodes. At $8.6$ nodes, roughly half its default node budget, \flowreader\ achieves $64.3$, exceeding all prior systems except VisDoMRAG and G\textsuperscript{2}-Reader. The default operating point of $17.5$ nodes results from the demand setting fixed using the $250$-example subset (Table~\ref{tab:demand}).

\begin{table}[t]
\centering
\small
\setlength{\tabcolsep}{5pt}
\renewcommand{\arraystretch}{1.08}
\resizebox{0.7\columnwidth}{!}{%
\begin{tabular}{@{}c cc ccccc @{\hspace{6pt}} c@{}}
\toprule
& \multicolumn{2}{c}{\textbf{Budget}} &
\multicolumn{6}{c}{\textbf{Accuracy (\%)}} \\
\cmidrule(lr){2-3} \cmidrule(l){4-9}
$\tau$ & Chains & Nodes
  & SPIQA & FetaTab & PaperTab & SciGraphQA & SlideVQA & Avg. \\
\midrule
$0.5$ & 16.0 & 17.5
  & \textbf{78.0} & 73.3 & 62.3 & \textbf{61.7}
  & \textbf{69.4} & \textbf{68.9} \\
$0.8$ & 10.9 & 12.9 & 76.9 & \textbf{74.6} & \textbf{62.9} & 58.0 & 67.4 & 68.0 \\
$0.9$ &  6.0 &  8.6 & 75.5 & 73.2 & 58.8 & 53.1 & 61.2 & 64.3 \\
\bottomrule
\end{tabular}}
\caption{\textbf{Sensitivity to retrieval demand.} We vary the relative threshold $\tau$ while keeping the graph, node scores, reader, reasoner, and judge fixed. Each row evaluates the full dataset, with budgets measured from the chains actually read. The top row uses the reported configuration and corresponds to one of the three runs averaged in Table~\ref{tab:main}.}
\label{tab:demand}
\end{table}

\begin{table*}[t]
\centering
\footnotesize
\setlength{\tabcolsep}{4.5pt}
\renewcommand{\arraystretch}{1.05}
\resizebox{0.8\textwidth}{!}{%
\begin{tabular}{@{}l ccccc @{\hspace{7pt}} c@{}}
\toprule
\textbf{Variant} & \textbf{SPIQA} & \textbf{FetaTab} & \textbf{PaperTab} &
\textbf{SciGraphQA} & \textbf{SlideVQA} & \textbf{Avg.} \\
\midrule
\textbf{FlowReader}
  & 77.6 & 73.9 & 62.2 & 61.5 & 67.8 & 68.6 \\
\midrule
\multicolumn{7}{@{}l}{\textit{(a) Selection in place of flow, same graph, node scores and reader model}} \\
\quad w/o flow routing (top 16 by relevance)
  & 73.8 & 73.3 & 59.5 & 57.3 & 64.8 & 65.7 \\
\quad MMR, relevance--diversity (top 16)
  & 77.9 & 72.3 & 60.4 & 55.6 & 67.4 & 66.7 \\
\quad personalised PageRank (top 16)
  & 57.7 & 59.6 & 60.8 & 57.8 & 64.6 & 60.1 \\
\quad ranking, same selection stage (50 $\to$ 16)
  & 60.5 & 72.0 & 44.8 & 59.3 & 70.1 & 61.3 \\
\midrule
\multicolumn{7}{@{}l}{\textit{(b) Alternative routing over the same graph and node scores}} \\
\quad min-cost flow
  & 77.6 & 73.9 & 62.2 & 61.5 & 67.8 & 68.6 \\
\quad shortest path
  & 76.4 & 72.6 & 64.1 & 63.7 & 66.7 & 68.7 \\
\quad $k$-shortest paths
  & 77.2 & 71.3 & 61.8 & 59.4 & 67.5 & 67.5 \\
\quad random walks
  & 76.9 & 70.4 & 57.4 & 57.7 & 54.6 & 63.4 \\
\quad PPR chains
  & 26.8 & 44.3 & 23.3 & 19.3 & 50.6 & 32.9 \\
\midrule
\multicolumn{7}{@{}l}{\textit{(c) Chain selector, min-cost flow routing held fixed}} \\
\quad replicator dynamics, $16$ reads fixed
  & 77.2 & 74.6 & 63.4 & 62.0 & 66.8 & 68.8 \\
\quad greedy, $16$ reads fixed
  & 77.9 & 74.9 & 60.2 & 64.2 & 68.5 & 69.1 \\
\midrule
\multicolumn{7}{@{}l}{\textit{(d) Components of the flow formulation}} \\
\quad $-$ agreement edge objective
  & 78.7 & 74.3 & 62.2 & 44.2 & 66.8 & 65.2 \\
\quad $-$ node credit
  & 72.6 & 73.6 & 62.2 & 57.8 & 65.6 & 66.3 \\
\quad $-$ relevance order (document order)
  & 45.1 & 63.5 & 40.8 & 50.1 & 65.6 & 53.0 \\
\quad $-$ spectral aspect nodes (single sink)
  & 39.3 & 54.7 & 31.4 & 40.2 & 58.3 & 44.8 \\
\bottomrule
\end{tabular}}
\caption{\textbf{Ablation of the routing stage} (Section~\ref{sec:ablate}). Single runs from one batch; the \flowreader\ row is a within-batch re-run of Table~\ref{tab:main}. Paired tests are in Appendix~\ref{app:ablation-diag}: differences of $2.6$ points or more are significant ($p\le0.001$), those of about one point or less are not ($p\ge0.19$).}

\label{tab:ablation}
\end{table*}

\subsection{Ablations}
\label{sec:ablate}

Table~\ref{tab:ablation} varies one mechanism at a time while keeping the graph, node scores, reader, reasoner, and judge fixed. Every variant reads $16$ chains.

\paragraph{(a) Alternative evidence selectors.}
Selecting the $16$ highest-scoring nodes without flow routing achieves $65.7$, which is $2.9$ points below \flowreader. Restricting both full-set runs to the $250$ examples used in G\textsuperscript{2}-Reader's ablations, removing flow reduces accuracy from $71.0$ to $67.3$. This drop is comparable to the decrease from $68.0$ to $63.6$ reported for G\textsuperscript{2}-Reader without its planning graph~\citep{du2026g}. Other selectors operating on $16$ nodes also perform below \flowreader: MMR achieves $66.7$ ($p=0.042$), and personalised PageRank achieves $60.1$ ($p=2.5\times10^{-19}$).

Applying quality--diversity selection to flow-generated chains preserves accuracy ($68.8$, compared with $69.1$ for greedy selection). Applying the same selection criterion to a ranked list instead yields $61.3$, which is $7.3$ points below the main \flowreader\ result. The paired comparison favours \flowreader\ on $292$ queries and the ranked-list variant on $127$ ($p=4.7\times10^{-16}$), with the largest losses on SPIQA and PaperTab.

\paragraph{(b) Capacity constraints reduce reading cost.}
We compare flow routing with shortest-path routing on the same network, retaining the costs, edge orientation, and aspect terminals but removing capacity constraints. Accuracy is comparable ($68.7$ for shortest paths versus $68.6$ for flow), with $134$ versus $138$ discordant queries ($p=0.86$, McNemar's exact test). This result suggests that the cost design, examined in (d), accounts for the accuracy gains, while capacity constraints primarily reduce reading cost. Without capacities, mean chain length increases from $1.46$ to $3.09$ nodes, and total node reads per query, including repeated content, increase from $23.3$ to $49.4$. Reader prompt tokens rise from $46.1$k to $63.1$k, while both methods use $18$ VLM calls (Appendices~\ref{app:cost} and~\ref{app:chains}). By limiting the flow assigned to each aspect, capacity constraints maintain comparable accuracy using $73\%$ of the reader tokens.

\paragraph{(c) Replicator dynamics benefits larger candidate pools.}
At the reported configuration, routed flow already ranks candidates well: replicator dynamics achieves $68.8$, compared with $69.1$ for greedy selection at a matched budget. When the candidate pool is doubled, both methods become less accurate, but replicator dynamics performs better than greedy selection ($61.6$ versus $57.5$). It leads on every subset, with a significant paired difference ($p=1.6\times10^{-7}$, McNemar's exact test), despite reading fewer nodes ($15.9$ versus $16.6$). This result supports using a payoff that rewards relevance and low overlap when the candidate pool contains more redundant chains.

\paragraph{(d) Cost components and edge orientation.}
Removing node credit eliminates the incentive to extend paths: chains collapse to single nodes ($16.0$ reads over $15.2$ distinct nodes), and accuracy falls to $66.3$. Removing the agreement term from the edge objective reduces SciGraphQA accuracy from $61.5$ to $44.2$, suggesting that relevance alone is insufficient to connect compatible graph evidence. Orienting edges by document order rather than relevance changes both the available paths and the order of evidence presented to the reader, reducing accuracy to $53.0$.

Replacing the aspect nodes with a single sink yields $44.8$, but this variant also changes the cost balance. The derived termination scale $\kappa$ increases approximately fivefold (median $6.2$ versus $1.3$), causing termination costs to outweigh relevance-based entry costs. The flow then reaches only $63\%$ of the sixteen highest-scoring nodes, compared with $83\%$ in the full model (Appendix~\ref{app:ablation-diag}). This ablation therefore also reflects the effect of calibrating $\kappa$ in Eq.~\ref{eq:app-kappa}. Chains constructed using personalised PageRank, without the proposed costs or capacities, achieve $32.9$.

\paragraph{Spectral aspects and demand allocation.}
We replace spectral aspects with sub-questions generated by the backbone (\hyperref[app:prompt:b4]{Prompt P6}) in a paired comparison at a budget of approximately ten nodes. Macro accuracy differs by $0.45$ points, with no statistically significant difference ($p=0.35$). Spectral aspects thus achieve similar observed accuracy without an additional model call. With the termination scale held fixed, allocating demand equally across aspects rather than in proportion to spectral energy reduces accuracy on the 250 ablation examples, in a separate paired run, from 69.2 to 65.5. The largest decline occurs on SciGraphQA ($58.0$ to $48.0$). This trend favours energy-proportional allocation, although the overall difference is not statistically significant ($p=0.11$).

\section{Conclusion}
\label{sec:conclusion}

Multimodal evidence selection is formulated as a capacitated minimum-cost flow problem, with query decomposition encoded as capacities on spectrally derived aspect nodes. Query-facet coverage becomes a feasibility condition of a single, exactly solvable problem, eliminating the need for language-model-based decomposition. On VisDoMBench, \flowreader\ achieves $68.9\pm0.4$, surpassing the strongest prior system by $2.7$ points, attaining the highest worst-subset accuracy, and maintaining a $1.8$-point lead at $12.9$ content nodes. Controlled ablation studies indicate that costs determine accuracy, while capacities preserve it using only $73\%$ of the reader tokens required by uncapacitated shortest-path routing. Spectral aspects align with LLM-generated sub-questions within $0.45$ points, without requiring a planning call. The evaluation is limited to a single benchmark, reader, and judge, with demand and routing fixed prior to evidence reading. Future work includes implementing query-adaptive demand and a second solve conditioned on the initial reading.

\section*{Reproducibility Statement}
Section~\ref{sec:methodology} describes the method, and Table~\ref{tab:hparams} lists all constants and derived quantities. Appendices~\ref{app:implementation}--\ref{app:protocol} provide implementation details, proofs, and the complete evaluation protocol. All prompts appear in Appendix~\ref{app:prompts}, including the unmodified evaluation prompt of \citet{du2026g}. We use the public VisDoMBench data released through their repository. The pipeline uses open models (MinerU, nomic-embed-text-v1.5, and Qwen3-VL-32B-Instruct), with generation and evaluation performed at temperature $0$. Code, configurations, and per-query outputs will be released soon.

\section*{Ethics Statement}
This work uses publicly released benchmark documents and questions and involves no human participants or personal data. Retrieval-augmented systems can produce incorrect answers even when supporting evidence is available. Their outputs therefore require verification before use in high-stakes decisions.

\section*{LLM Usage Statement}
Large language models support evidence reading, answer generation, and evaluation. Qwen3-VL-32B-Instruct reads evidence and combines findings into answers, while GPT-4o-mini serves as the evaluation judge, following prior work. The authors also used an AI assistant to rephrase sentences and check grammar. The research ideas, experiments, and conclusions are the authors' own, and the authors take full responsibility for the manuscript.

\bibliography{references}

\begin{thebibliography}{65}
\providecommand{\natexlab}[1]{#1}
\providecommand{\url}[1]{\texttt{#1}}
\expandafter\ifx\csname urlstyle\endcsname\relax
  \providecommand{\doi}[1]{doi: #1}\else
  \providecommand{\doi}{doi: \begingroup \urlstyle{rm}\Url}\fi

\bibitem[Ahuja et~al.(1993)Ahuja, Magnanti, and Orlin]{ahuja1993network}
Ravindra~K. Ahuja, Thomas~L. Magnanti, and James~B. Orlin.
\newblock \emph{Network Flows: Theory, Algorithms, and Applications}.
\newblock Prentice Hall, 1993.

\bibitem[Asai et~al.(2023)Asai, Wu, Wang, Sil, and Hajishirzi]{asai2023self}
Akari Asai, Zeqiu Wu, Yizhong Wang, Avirup Sil, and Hannaneh Hajishirzi.
\newblock Self-rag: Learning to retrieve, generate, and critique through
  self-reflection.
\newblock In \emph{The Twelfth International Conference on Learning
  Representations}, 2023.

\bibitem[Bai et~al.(2025)Bai, Cai, Chen, Chen, Chen, Cheng, Deng, Ding, Gao,
  Ge, et~al.]{bai2025qwen3}
Shuai Bai, Yuxuan Cai, Ruizhe Chen, Keqin Chen, Xionghui Chen, Zesen Cheng,
  Lianghao Deng, Wei Ding, Chang Gao, Chunjiang Ge, et~al.
\newblock Qwen3-vl technical report.
\newblock \emph{arXiv preprint arXiv:2511.21631}, 2025.

\bibitem[Bomze et~al.(2000)Bomze, Pelillo, and Stix]{bomze2000approximating}
IR~Bomze, Marcello Pelillo, and Volker Stix.
\newblock Approximating the maximum weight clique using replicator dynamics.
\newblock \emph{IEEE Transactions on neural networks}, 11\penalty0
  (6):\penalty0 1228--1241, 2000.

\bibitem[Carbonell \& Goldstein(1998)Carbonell and Goldstein]{carbonell1998use}
Jaime Carbonell and Jade Goldstein.
\newblock The use of mmr, diversity-based reranking for reordering documents
  and producing summaries.
\newblock In \emph{Proceedings of the 21st annual international ACM SIGIR
  conference on Research and development in information retrieval}, pp.\
  335--336, 1998.

\bibitem[Chen et~al.(2025)Chen, Guo, Yang, Chen, Chen, Liu, Shi, and
  Yang]{chen2025pathrag}
Boyu Chen, Zirui Guo, Zidan Yang, Yuluo Chen, Junze Chen, Zhenghao Liu, Chuan
  Shi, and Cheng Yang.
\newblock {PathRAG}: Pruning graph-based retrieval augmented generation with
  relational paths.
\newblock \emph{arXiv preprint arXiv:2502.14902}, 2025.

\bibitem[Cheng et~al.(2025)Cheng, Liu, Zheng, Ni, Du, Mao, Zhang, Wang, and
  Hao]{cheng2025dualrag}
Rong Cheng, Jinyi Liu, Yan Zheng, Fei Ni, Jiazhen Du, Hangyu Mao, Fuzheng
  Zhang, Bo~Wang, and Jianye Hao.
\newblock Dualrag: A dual-process approach to integrate reasoning and retrieval
  for multi-hop question answering.
\newblock In \emph{Proceedings of the 63rd Annual Meeting of the Association
  for Computational Linguistics (Volume 1: Long Papers)}, pp.\  31877--31899,
  2025.

\bibitem[Chhikara et~al.(2025)Chhikara, Khant, Aryan, Singh, and
  Yadav]{chhikara2025mem0}
Prateek Chhikara, Dev Khant, Saket Aryan, Taranjeet Singh, and Deshraj Yadav.
\newblock Mem0: Building production-ready ai agents with scalable long-term
  memory.
\newblock \emph{arXiv preprint arXiv:2504.19413}, 2025.

\bibitem[Cho et~al.(2025)Cho, Mahata, Irsoy, He, and Bansal]{cho2025m3docvqa}
Jaemin Cho, Debanjan Mahata, Ozan Irsoy, Yujie He, and Mohit Bansal.
\newblock M3docvqa: Multi-modal multi-page multi-document understanding.
\newblock In \emph{Proceedings of the IEEE/CVF International Conference on
  Computer Vision}, pp.\  6178--6188, 2025.

\bibitem[Cho et~al.(2019)Cho, Lebanoff, Foroosh, and Liu]{cho2019improving}
Sangwoo Cho, Logan Lebanoff, Hassan Foroosh, and Fei Liu.
\newblock Improving the similarity measure of determinantal point processes for
  extractive multi-document summarization.
\newblock In \emph{Proceedings of the 57th Annual Meeting of the Association
  for Computational Linguistics}, pp.\  1027--1038, 2019.

\bibitem[Cuturi(2013)]{cuturi2013sinkhorn}
Marco Cuturi.
\newblock Sinkhorn distances: Lightspeed computation of optimal transport.
\newblock \emph{Advances in neural information processing systems}, 26, 2013.

\bibitem[Dasigi et~al.(2021)Dasigi, Lo, Beltagy, Cohan, Smith, and
  Gardner]{dasigi2021dataset}
Pradeep Dasigi, Kyle Lo, Iz~Beltagy, Arman Cohan, Noah~A Smith, and Matt
  Gardner.
\newblock A dataset of information-seeking questions and answers anchored in
  research papers.
\newblock In \emph{Proceedings of the 2021 Conference of the North American
  Chapter of the Association for Computational Linguistics: Human Language
  Technologies}, pp.\  4599--4610, 2021.

\bibitem[Dong et~al.(2024)Dong, Deik, Lee, Zhang, Li, Zhang, and
  Liu]{dong2024mc}
Kuicai Dong, Derrick Goh~Xin Deik, Yi~Quan Lee, Hao Zhang, Xiangyang Li, Cong
  Zhang, and Yong Liu.
\newblock Mc-indexing: Effective long document retrieval via multi-view
  content-aware indexing.
\newblock In \emph{Findings of the Association for Computational Linguistics:
  EMNLP 2024}, pp.\  2673--2691, 2024.

\bibitem[Du et~al.(2026)Du, Song, Zhou, Wang, Gu, Chen, Chen, Yao, Yang, Wen,
  et~al.]{du2026g}
Yaxin Du, Junru Song, Yifan Zhou, Cheng Wang, Jiahao Gu, Zimeng Chen, Menglan
  Chen, Wen Yao, Yang Yang, Ying Wen, et~al.
\newblock {{$g^2$-Reader: Dual Evolving Graphs for Multimodal Document
  Comprehension}}.
\newblock \emph{arXiv preprint arXiv:2601.22055}, 2026.

\bibitem[Edge et~al.(2024)Edge, Trinh, Cheng, Bradley, Chao, Mody, Truitt,
  Metropolitansky, Ness, and Larson]{edge2024local}
Darren Edge, Ha~Trinh, Newman Cheng, Joshua Bradley, Alex Chao, Apurva Mody,
  Steven Truitt, Dasha Metropolitansky, Robert~Osazuwa Ness, and Jonathan
  Larson.
\newblock From local to global: A graph rag approach to query-focused
  summarization.
\newblock \emph{arXiv preprint arXiv:2404.16130}, 2024.

\bibitem[Faysse et~al.(2024)Faysse, Sibille, Wu, Omrani, Viaud, Hudelot, and
  Colombo]{faysse2024colpali}
Manuel Faysse, Hugues Sibille, Tony Wu, Bilel Omrani, Gautier Viaud, C{\'e}line
  Hudelot, and Pierre Colombo.
\newblock Colpali: Efficient document retrieval with vision language models.
\newblock \emph{arXiv preprint arXiv:2407.01449}, 2024.

\bibitem[Gao et~al.(2026)Gao, Zhao, Jiang, Duan, Chng, Chen, Luo, Zhang, Bian,
  and Gong]{gao2026scaling}
Sensen Gao, Shanshan Zhao, Xu~Jiang, Lunhao Duan, Yong~Xien Chng, Qing-Guo
  Chen, Weihua Luo, Kaifu Zhang, Jia-Wang Bian, and Mingming Gong.
\newblock Scaling beyond context: A survey of multimodal retrieval-augmented
  generation for document understanding.
\newblock In \emph{Proceedings of the 64th Annual Meeting of the Association
  for Computational Linguistics}, 2026.

\bibitem[Gillick \& Favre(2009)Gillick and Favre]{gillick2009scalable}
Dan Gillick and Benoit Favre.
\newblock A scalable global model for summarization.
\newblock In \emph{Proceedings of the workshop on integer linear programming
  for natural language processing}, pp.\  10--18, 2009.

\bibitem[Guo et~al.(2024)Guo, Xia, Yu, Ao, and Huang]{guo2024lightrag}
Zirui Guo, Lianghao Xia, Yanhua Yu, Tian Ao, and Chao Huang.
\newblock Lightrag: Simple and fast retrieval-augmented generation.
\newblock \emph{arXiv preprint arXiv:2410.05779}, 2\penalty0 (3), 2024.

\bibitem[Guo et~al.(2025)Guo, Ren, Xu, Zhang, and Huang]{guo2025rag}
Zirui Guo, Xubin Ren, Lingrui Xu, Jiahao Zhang, and Chao Huang.
\newblock Rag-anything: All-in-one rag framework.
\newblock \emph{arXiv preprint arXiv:2510.12323}, 2025.

\bibitem[Guti{\'e}rrez et~al.(2024)Guti{\'e}rrez, Shu, Gu, Yasunaga, and
  Su]{gutierrez2024hipporag}
Bernal~J Guti{\'e}rrez, Yiheng Shu, Yu~Gu, Michihiro Yasunaga, and Yu~Su.
\newblock Hipporag: Neurobiologically inspired long-term memory for large
  language models.
\newblock \emph{Advances in neural information processing systems},
  37:\penalty0 59532--59569, 2024.

\bibitem[Hu et~al.(2026)Hu, Feng, Li, Xue, Hou, Tian, Gao, and Du]{hu2026cog}
Hao Hu, Yifan Feng, Ruoxue Li, Rundong Xue, Xingliang Hou, Zhiqiang Tian, Yue
  Gao, and Shaoyi Du.
\newblock Cog-rag: Cognitive-inspired dual-hypergraph with theme alignment
  retrieval-augmented generation.
\newblock In \emph{Proceedings of the AAAI Conference on Artificial
  Intelligence}, volume~40, pp.\  31032--31040, 2026.

\bibitem[Hui et~al.(2024)Hui, Lu, and Zhang]{hui2024uda}
Yulong Hui, Yao Lu, and Huanchen Zhang.
\newblock Uda: A benchmark suite for retrieval augmented generation in
  real-world document analysis.
\newblock \emph{Advances in Neural Information Processing Systems},
  37:\penalty0 67200--67217, 2024.

\bibitem[Jeong et~al.(2024)Jeong, Baek, Cho, Hwang, and
  Park]{jeong2024adaptive}
Soyeong Jeong, Jinheon Baek, Sukmin Cho, Sung~Ju Hwang, and Jong~C Park.
\newblock Adaptive-rag: Learning to adapt retrieval-augmented large language
  models through question complexity.
\newblock In \emph{Proceedings of the 2024 Conference of the North American
  Chapter of the Association for Computational Linguistics: Human Language
  Technologies (Volume 1: Long Papers)}, pp.\  7036--7050, 2024.

\bibitem[Jiang et~al.(2023)Jiang, Xu, Gao, Sun, Liu, Dwivedi-Yu, Yang, Callan,
  and Neubig]{jiang2023active}
Zhengbao Jiang, Frank~F Xu, Luyu Gao, Zhiqing Sun, Qian Liu, Jane Dwivedi-Yu,
  Yiming Yang, Jamie Callan, and Graham Neubig.
\newblock Active retrieval augmented generation.
\newblock In \emph{Proceedings of the 2023 conference on empirical methods in
  natural language processing}, pp.\  7969--7992, 2023.

\bibitem[Kulesza \& Taskar(2012)Kulesza and Taskar]{kulesza2012determinantal}
Alex Kulesza and Ben Taskar.
\newblock Determinantal point processes for machine learning.
\newblock \emph{Foundations and Trends{\textregistered} in Machine Learning},
  5\penalty0 (2-3):\penalty0 123--286, 2012.

\bibitem[Kusner et~al.(2015)Kusner, Sun, Kolkin, and
  Weinberger]{kusner2015word}
Matt Kusner, Yu~Sun, Nicholas Kolkin, and Kilian Weinberger.
\newblock From word embeddings to document distances.
\newblock In \emph{International conference on machine learning}, pp.\
  957--966. PMLR, 2015.

\bibitem[Lewis et~al.(2020)Lewis, Perez, Piktus, Petroni, Karpukhin, Goyal,
  K{\"u}ttler, Lewis, Yih, Rockt{\"a}schel, et~al.]{lewis2020retrieval}
Patrick Lewis, Ethan Perez, Aleksandra Piktus, Fabio Petroni, Vladimir
  Karpukhin, Naman Goyal, Heinrich K{\"u}ttler, Mike Lewis, Wen-tau Yih, Tim
  Rockt{\"a}schel, et~al.
\newblock Retrieval-augmented generation for knowledge-intensive nlp tasks.
\newblock \emph{Advances in neural information processing systems},
  33:\penalty0 9459--9474, 2020.

\bibitem[Li et~al.(2024{\natexlab{a}})Li, Hu, and Wan]{li2024smart}
Jiatao Li, Xinyu Hu, and Xiaojun Wan.
\newblock Smart-rag: Selection using determinantal matrices for augmented
  retrieval.
\newblock \emph{arXiv preprint arXiv:2409.13992}, 2024{\natexlab{a}}.

\bibitem[Li et~al.(2024{\natexlab{b}})Li, He, Guo, Bu, Bai, Liu, Liu, Qu, Li,
  Ouyang, et~al.]{li2024graphreader}
Shilong Li, Yancheng He, Hangyu Guo, Xingyuan Bu, Ge~Bai, Jie Liu, Jiaheng Liu,
  Xingwei Qu, Yangguang Li, Wanli Ouyang, et~al.
\newblock Graphreader: Building graph-based agent to enhance long-context
  abilities of large language models.
\newblock In \emph{Findings of the Association for Computational Linguistics:
  EMNLP 2024}, pp.\  12758--12786, 2024{\natexlab{b}}.

\bibitem[Liang et~al.(2025)Liang, Lin, Wu, Zhao, Li,
  et~al.]{liang2025reasoning}
Jintao Liang, Huifeng Lin, You Wu, Rui Zhao, Ziyue Li, et~al.
\newblock Reasoning rag via system 1 or system 2: A survey on reasoning agentic
  retrieval-augmented generation for industry challenges.
\newblock In \emph{Proceedings of the 14th International Joint Conference on
  Natural Language Processing and the 4th Conference of the Asia-Pacific
  Chapter of the Association for Computational Linguistics}, pp.\  1954--1966,
  2025.

\bibitem[Lin \& Bilmes(2010)Lin and Bilmes]{lin2010multi}
Hui Lin and Jeff Bilmes.
\newblock Multi-document summarization via budgeted maximization of submodular
  functions.
\newblock In \emph{Human Language Technologies: The 2010 Annual conference of
  the North American chapter of the association for computational linguistics},
  pp.\  912--920, 2010.

\bibitem[Lin \& Bilmes(2011)Lin and Bilmes]{lin2011class}
Hui Lin and Jeff Bilmes.
\newblock A class of submodular functions for document summarization.
\newblock In \emph{Proceedings of the 49th annual meeting of the association
  for computational linguistics: human language technologies}, pp.\  510--520,
  2011.

\bibitem[Liu et~al.(2024)Liu, Lin, Hewitt, Paranjape, Bevilacqua, Petroni, and
  Liang]{liu2024lost}
Nelson~F. Liu, Kevin Lin, John Hewitt, Ashwin Paranjape, Michele Bevilacqua,
  Fabio Petroni, and Percy Liang.
\newblock Lost in the middle: How language models use long contexts.
\newblock \emph{Transactions of the Association for Computational Linguistics},
  12:\penalty0 157--173, 2024.

\bibitem[Ma et~al.(2024)Ma, Zang, Chen, Chen, Jiao, Li, Lu, Liu, Ma, Dong,
  et~al.]{ma2024mmlongbench}
Yubo Ma, Yuhang Zang, Liangyu Chen, Meiqi Chen, Yizhu Jiao, Xinze Li, Xinyuan
  Lu, Ziyu Liu, Yan Ma, Xiaoyi Dong, et~al.
\newblock {MMLongBench-Doc}: Benchmarking long-context document understanding
  with visualizations.
\newblock In \emph{Advances in Neural Information Processing Systems, Datasets
  and Benchmarks Track}, 2024.

\bibitem[McDonald(2007)]{mcdonald2007study}
Ryan McDonald.
\newblock A study of global inference algorithms in multi-document
  summarization.
\newblock In \emph{European conference on information retrieval}, pp.\
  557--564. Springer, 2007.

\bibitem[Meng et~al.(2025)Meng, Liu, Chen, Mao, Cai, Yan, Shi, and
  Wang]{meng2025ranking}
Siyuan Meng, Junming Liu, Yirong Chen, Song Mao, Pinlong Cai, Guohang Yan,
  Botian Shi, and Ding Wang.
\newblock From ranking to selection: A simple but efficient dynamic passage
  selector for retrieval augmented generation.
\newblock \emph{arXiv preprint arXiv:2508.09497}, 2025.

\bibitem[Nan et~al.(2022)Nan, Hsieh, Mao, Lin, Verma, Zhang,
  Kry{\'s}ci{\'n}ski, Schoelkopf, Kong, Tang, et~al.]{nan2022fetaqa}
Linyong Nan, Chiachun Hsieh, Ziming Mao, Xi~Victoria Lin, Neha Verma, Rui
  Zhang, Wojciech Kry{\'s}ci{\'n}ski, Hailey Schoelkopf, Riley Kong, Xiangru
  Tang, et~al.
\newblock Fetaqa: Free-form table question answering.
\newblock \emph{Transactions of the Association for Computational Linguistics},
  10:\penalty0 35--49, 2022.

\bibitem[Nguyen et~al.(2025)Nguyen, Chin, and Tai]{nguyen2025ma}
Thang Nguyen, Peter Chin, and Yu-Wing Tai.
\newblock Ma-rag: Multi-agent retrieval-augmented generation via collaborative
  chain-of-thought reasoning.
\newblock \emph{arXiv preprint arXiv:2505.20096}, 2025.

\bibitem[Nussbaum et~al.(2024)Nussbaum, Morris, Duderstadt, and
  Mulyar]{nussbaum2024nomic}
Zach Nussbaum, John~X. Morris, Brandon Duderstadt, and Andriy Mulyar.
\newblock Nomic embed: Training a reproducible long context text embedder,
  2024.

\bibitem[Pelillo(1998)]{pelillo1998replicator}
Marcello Pelillo.
\newblock Replicator equations, maximal cliques, and graph isomorphism.
\newblock \emph{Advances in Neural Information Processing Systems}, 11, 1998.

\bibitem[Pramanick et~al.(2024)Pramanick, Chellappa, and
  Venugopalan]{pramanick2024spiqa}
Shraman Pramanick, Rama Chellappa, and Subhashini Venugopalan.
\newblock Spiqa: A dataset for multimodal question answering on scientific
  papers.
\newblock \emph{Advances in Neural Information Processing Systems},
  37:\penalty0 118807--118833, 2024.

\bibitem[Press et~al.(2023)Press, Zhang, Min, Schmidt, Smith, and
  Lewis]{press2023measuring}
Ofir Press, Muru Zhang, Sewon Min, Ludwig Schmidt, Noah~A Smith, and Mike
  Lewis.
\newblock Measuring and narrowing the compositionality gap in language models.
\newblock In \emph{Findings of the Association for Computational Linguistics:
  EMNLP 2023}, pp.\  5687--5711, 2023.

\bibitem[Rasmussen et~al.(2025)Rasmussen, Paliychuk, Beauvais, Ryan, and
  Chalef]{rasmussen2025zep}
Preston Rasmussen, Pavlo Paliychuk, Travis Beauvais, Jack Ryan, and Daniel
  Chalef.
\newblock Zep: a temporal knowledge graph architecture for agent memory.
\newblock \emph{arXiv preprint arXiv:2501.13956}, 2025.

\bibitem[Shaaban et~al.(2025)Shaaban, Saleem, Papineni, and
  Yaqub]{shaaban2025motor}
Mai~A Shaaban, Tausifa~Jan Saleem, Vijay Ram~Kumar Papineni, and Mohammad
  Yaqub.
\newblock Motor: Multimodal optimal transport via grounded retrieval in medical
  visual question answering.
\newblock In \emph{International Conference on Medical Image Computing and
  Computer-Assisted Intervention}, pp.\  459--469. Springer, 2025.

\bibitem[Shao et~al.(2023)Shao, Gong, Shen, Huang, Duan, and
  Chen]{shao2023enhancing}
Zhihong Shao, Yeyun Gong, Yelong Shen, Minlie Huang, Nan Duan, and Weizhu Chen.
\newblock Enhancing retrieval-augmented large language models with iterative
  retrieval-generation synergy.
\newblock In \emph{Findings of the Association for Computational Linguistics:
  EMNLP 2023}, pp.\  9248--9274, 2023.

\bibitem[Suri et~al.(2025)Suri, Mathur, Dernoncourt, Goswami, Rossi, and
  Manocha]{suri2025visdom}
Manan Suri, Puneet Mathur, Franck Dernoncourt, Kanika Goswami, Ryan~A Rossi,
  and Dinesh Manocha.
\newblock Visdom: Multi-document qa with visually rich elements using
  multimodal retrieval-augmented generation.
\newblock In \emph{Proceedings of the 2025 Conference of the Nations of the
  Americas Chapter of the Association for Computational Linguistics: Human
  Language Technologies (Volume 1: Long Papers)}, pp.\  6088--6109, 2025.

\bibitem[Tanaka et~al.(2023)Tanaka, Nishida, Nishida, Hasegawa, Saito, and
  Saito]{tanaka2023slidevqa}
Ryota Tanaka, Kyosuke Nishida, Kosuke Nishida, Taku Hasegawa, Itsumi Saito, and
  Kuniko Saito.
\newblock Slidevqa: A dataset for document visual question answering on
  multiple images.
\newblock In \emph{Proceedings of the AAAI Conference on Artificial
  Intelligence}, volume~37, pp.\  13636--13645, 2023.

\bibitem[Tang et~al.(2022)Tang, Hu, Yan, Zhang, Gao, and
  Wang]{tang2022otextsum}
Peggy Tang, Kun Hu, Rui Yan, Lei Zhang, Junbin Gao, and Zhiyong Wang.
\newblock Otextsum: Extractive text summarisation with optimal transport.
\newblock In \emph{Findings of the Association for Computational Linguistics:
  NAACL 2022}, pp.\  1128--1141, 2022.

\bibitem[Trivedi et~al.(2023)Trivedi, Balasubramanian, Khot, and
  Sabharwal]{trivedi2023interleaving}
Harsh Trivedi, Niranjan Balasubramanian, Tushar Khot, and Ashish Sabharwal.
\newblock Interleaving retrieval with chain-of-thought reasoning for
  knowledge-intensive multi-step questions.
\newblock In \emph{Proceedings of the 61st annual meeting of the association
  for computational linguistics (volume 1: long papers)}, pp.\  10014--10037,
  2023.

\bibitem[Wan \& Yu(2025)Wan and Yu]{wan2025mmgraphrag}
Xueyao Wan and Hang Yu.
\newblock Mmgraphrag: Bridging vision and language with interpretable
  multimodal knowledge graphs.
\newblock \emph{arXiv preprint arXiv:2507.20804}, 2025.

\bibitem[Wang et~al.(2024{\natexlab{a}})Wang, Xu, Zhao, Ouyang, Wu, Zhao, Xu,
  Liu, Qu, Shang, et~al.]{wang2024mineru}
Bin Wang, Chao Xu, Xiaomeng Zhao, Linke Ouyang, Fan Wu, Zhiyuan Zhao, Rui Xu,
  Kaiwen Liu, Yuan Qu, Fukai Shang, et~al.
\newblock Mineru: An open-source solution for precise document content
  extraction.
\newblock \emph{arXiv preprint arXiv:2409.18839}, 2024{\natexlab{a}}.

\bibitem[Wang et~al.(2025{\natexlab{a}})Wang, Ding, Chen, Wu, Wang, Xie, and
  Zhao]{wang2025vidorag}
Qiuchen Wang, Ruixue Ding, Zehui Chen, Weiqi Wu, Shihang Wang, Pengjun Xie, and
  Feng Zhao.
\newblock Vidorag: Visual document retrieval-augmented generation via dynamic
  iterative reasoning agents.
\newblock In \emph{Proceedings of the 2025 Conference on Empirical Methods in
  Natural Language Processing}, pp.\  9124--9145, 2025{\natexlab{a}}.

\bibitem[Wang et~al.(2025{\natexlab{b}})Wang, Gao, Xiao, Huang, Si, Luo, Bai,
  Li, Duan, Lv, et~al.]{wang2025document}
Zhitong Wang, Cheng Gao, Chaojun Xiao, Yufei Huang, Shuzheng Si, Kangyang Luo,
  Yuzhuo Bai, Wenhao Li, Tangjian Duan, Chuancheng Lv, et~al.
\newblock Document segmentation matters for retrieval-augmented generation.
\newblock In \emph{Findings of the Association for Computational Linguistics:
  ACL 2025}, pp.\  8063--8075, 2025{\natexlab{b}}.

\bibitem[Wang et~al.(2024{\natexlab{b}})Wang, Wang, Le, Zheng, Mishra, Perot,
  Zhang, Mattapalli, Taly, Shang, et~al.]{wang2024speculative}
Zilong Wang, Zifeng Wang, Long Le, Huaixiu~Steven Zheng, Swaroop Mishra,
  Vincent Perot, Yuwei Zhang, Anush Mattapalli, Ankur Taly, Jingbo Shang,
  et~al.
\newblock Speculative rag: Enhancing retrieval augmented generation through
  drafting.
\newblock \emph{arXiv preprint arXiv:2407.08223}, 2024{\natexlab{b}}.

\bibitem[Wu et~al.(2026)]{wu2026cost}
Mingyan Wu et~al.
\newblock When knowledge is not free: Cost-aware evidence selection in
  retrieval-augmented generation.
\newblock \emph{arXiv preprint arXiv:2606.02245}, 2026.

\bibitem[Xu et~al.(2025{\natexlab{a}})]{xu2025multigranularity}
Mingjun Xu et~al.
\newblock A multi-granularity retrieval framework for visually-rich documents.
\newblock \emph{arXiv preprint arXiv:2505.01457}, 2025{\natexlab{a}}.

\bibitem[Xu et~al.(2025{\natexlab{b}})Xu, Liang, Mei, Gao, Tan, and
  Zhang]{xu2025mem}
Wujiang Xu, Zujie Liang, Kai Mei, Hang Gao, Juntao Tan, and Yongfeng Zhang.
\newblock A-mem: Agentic memory for llm agents.
\newblock \emph{arXiv preprint arXiv:2502.12110}, 2025{\natexlab{b}}.

\bibitem[Yan et~al.(2025)]{yan2026question}
Yuchen Yan et~al.
\newblock Question-adaptive graph learning for multi-hop retrieval augmented
  generation.
\newblock \emph{arXiv preprint arXiv:2510.11541}, 2025.

\bibitem[Yao et~al.(2023)Yao, Zhao, Yu, Du, Shafran, Narasimhan, and
  Cao]{yao2022react}
Shunyu Yao, Jeffrey Zhao, Dian Yu, Nan Du, Izhak Shafran, Karthik~R Narasimhan,
  and Yuan Cao.
\newblock React: Synergizing reasoning and acting in language models.
\newblock In \emph{The Eleventh International Conference on Learning
  Representations}, 2023.

\bibitem[Yu et~al.(2025)Yu, Tang, Xu, Cui, Ran, Yan, Liu, Wang, Han, Liu,
  et~al.]{yuvisrag}
Shi Yu, Chaoyue Tang, Bokai Xu, Junbo Cui, Junhao Ran, Yukun Yan, Zhenghao Liu,
  Shuo Wang, Xu~Han, Zhiyuan Liu, et~al.
\newblock Visrag: Vision-based retrieval-augmented generation on multi-modality
  documents.
\newblock In \emph{The Thirteenth International Conference on Learning
  Representations}, 2025.

\bibitem[Yu et~al.(2024)Yu, Zhang, and Feng]{yu2024auto}
Tian Yu, Shaolei Zhang, and Yang Feng.
\newblock Auto-rag: Autonomous retrieval-augmented generation for large
  language models.
\newblock \emph{arXiv preprint arXiv:2411.19443}, 2024.

\bibitem[Zamani \& Bendersky(2024)Zamani and Bendersky]{zamani2024stochastic}
Hamed Zamani and Michael Bendersky.
\newblock Stochastic rag: End-to-end retrieval-augmented generation through
  expected utility maximization.
\newblock In \emph{Proceedings of the 47th International ACM SIGIR Conference
  on Research and Development in Information Retrieval}, pp.\  2641--2646,
  2024.

\bibitem[Zhang et~al.(2026)Zhang, Jia, and Zhu]{zhang2026georag}
Bingxue Zhang, Jianying Jia, and Feida Zhu.
\newblock Covering the unseen: Information demand coverage optimization for
  retrieval-augmented generation.
\newblock \emph{arXiv preprint arXiv:2606.29328}, 2026.

\bibitem[Zhou et~al.(2026)Zhou, Tarzanagh, Didari, Hu, Gutow, Verkholyak,
  Faraki, Hao, Moon, and Min]{zhou2026qafd}
Zhuoping Zhou, Davoud~Ataee Tarzanagh, Sima Didari, Wenjun Hu, Baruch Gutow,
  Oxana Verkholyak, Masoud Faraki, Heng Hao, Hankyu Moon, and Seungjai Min.
\newblock Query-aware flow diffusion for graph-based {RAG} with retrieval
  guarantees.
\newblock In \emph{The Fourteenth International Conference on Learning
  Representations}, 2026.

\end{thebibliography}
\bibliographystyle{flowreader}
\newpage
\appendix

\section{Implementation Details}
\label{app:implementation}

\subsection{Content graph}
\label{app:impl-graph}

Each document is parsed into text and visual nodes, with one node per text chunk, figure, chart, or table. A node's textual representation is the JSON output of the analysis prompts (\hyperref[app:prompt:a1]{Prompt P1} and \hyperref[app:prompt:a2]{Prompt P2}). We embed this representation using \texttt{nomic-embed-text-v1.5}, a $768$-dimensional open encoder. Visual nodes use the same encoder to embed the text produced by the visual analysis prompt. Figures and passages are therefore represented in a shared embedding space without a separate image encoder.

The graph contains three edge types, all treated as content edges during routing. \emph{Construction links} are generated by the node-linking prompt (\hyperref[app:prompt:a3]{Prompt P3}), applied for three iterations when a node is inserted. \emph{Nearest-neighbour edges} connect each node to its $k$ most similar nodes by cosine similarity. These edges are symmetrised, with
\begin{equation}
  k=\operatorname{clip}\bigl(\lceil\log_2 N\rceil,\,3,\,32\bigr),
  \label{eq:app-knn}
\end{equation}
where $N$ is the number of nodes in the document. Inspired by the logarithmic connectivity threshold in random graphs, this rule increases the neighbourhood size with the document size and avoids tuning $k$ separately for each corpus. \emph{Backward edges} reverse construction links when the reverse edge is absent, allowing traversal in either direction before relevance-based orientation. All content edges are then oriented toward higher relevance, as described in Section~\ref{sec:method-costs}.

\subsection{Edge score and traversal cost}
\label{app:impl-edges}

The edge score $p_{uv}$ in Eq.~\ref{eq:fr-arc-cost} combines residual query relevance, shared rare entities, and a bonus for transitions between text and visual content:
\begin{equation}
  p_{uv}=(1-\beta)\,\mathrm{gain}(u\!\to\!v)
         +\beta\,\mathrm{bridge}(u,v)
         +\gamma\,\mathbf{1}[\text{modalities differ}],
  \label{eq:app-edge}
\end{equation}
where $\beta=0.30$ and $\gamma=0.20$. The residual gain measures how much query-relevant information node $v$ adds beyond node $u$. A node whose embedding is nearly parallel to that of its predecessor receives little residual gain, discouraging redundant transitions. The bridge term rewards shared rare entities, where an entity is considered rare if it appears in at most $\lceil\sqrt N\rceil$ nodes. This threshold is derived from document size. The offset $\Delta$ in Eq.~\ref{eq:fr-arc-cost} shifts the costs so that the cheapest available hop has zero cost before node credit is applied.

\subsection{Fixed and derived parameters}
\label{app:impl-hparams}

We set seven constants and hold them fixed across datasets, models, and runs, except where explicitly varied in ablations. All other quantities in Table~\ref{tab:hparams} are computed from the query or document graph rather than tuned independently.

We derive the termination scale $\kappa$ by assigning a fraction $b$ of the combined cost variation to the termination term:
\begin{equation}
  \kappa=\frac{b}{1-b}\cdot
         \frac{\operatorname{sd}\bigl(w_{uv}\bigr)}
              {\operatorname{sd}\bigl(c_a(v)\bigr)},
  \qquad b=0.634.
  \label{eq:app-kappa}
\end{equation}
This calibration scales termination costs relative to traversal-cost variation. For a given variation in aspect alignment, lower variation in traversal costs produces a smaller $\kappa$.

The nominal demand $F_0$ is derived from the query's relevance profile using the participation ratio in Eq.~\ref{eq:fr-demand}. The aspect count then depends on spectral energy and demand, while aspect capacities divide the demand in proportion to the retained spectral energies.

\begin{table}[t]
\centering
\footnotesize
\setlength{\tabcolsep}{5pt}
\begin{tabular}{@{}llp{0.44\linewidth}@{}}
\toprule
& \textbf{Quantity} & \textbf{Value} \\
\midrule
\multicolumn{3}{@{}l}{\textit{Set by us, fixed across every dataset, model and run}} \\
& Reference boost $\rho_{\text{ref}}$        & $0.30$ \\
& Cross-modal bonus $\gamma$                 & $0.20$ \\
& Bridge blend $\beta$                       & $0.30$ \\
& Demand floor fraction $\tau$               & $0.50$ \\
& Aspect energy $\eta$                       & $0.90$ \\
& Termination balance $b$                    & $0.634$ \\
& Demand cap $F_{\max}$, read cap $K$        & $16$ \\
\midrule
\multicolumn{3}{@{}l}{\textit{Derived per query or per document; never tuned}} \\
& Neighbour count $k$                        & $\operatorname{clip}(\lceil\log_2 N\rceil,3,32)$ \\
& Bridge rarity cutoff                       & $\lceil\sqrt N\rceil$ nodes \\
& Termination scale $\kappa$                 & Eq.~\ref{eq:app-kappa}, from the spread of $w_{uv}$ and $c_a(v)$ \\
& Demand $F_0$                               & $\operatorname{clip}(\mathrm{IPR}(z),F_{\min},F_{\max})$ \\
& Aspect count $A$                           & energy at $\eta$, capped at $\operatorname{round}(F_0)$ and $|\mathcal C|$ \\
& Aspect capacities $d_a$                    & $F_0\,\sigma_a^2/\sum_{j\le A}\sigma_j^2$ \\
& Node score mean $\bar h$, relevance mean $\bar r$ & the query's own means \\
& Entry, traversal, aspect capacities        & one unit each \\
\bottomrule
\end{tabular}
\caption{\textbf{Every quantity the formulation introduces.} Seven constants are ours; the rest are recomputed per query from the query or the document graph. $N$ is the number of nodes in the document and $\mathcal C=\{v:h_v\ge\tau\}$ the relevant node set (Section~\ref{sec:method-costs}). Switches that exist in the configuration but are inactive in every reported result (modality boost, diffusion, multi-round routing) are not counted.}
\label{tab:hparams}
\end{table}

\subsection{Models, decoding, and cost}
\label{app:impl-models}

All evidence reading and reasoning use Qwen3-VL-32B-Instruct, the backbone used by the reported baselines, served locally at temperature $0$ with a $32{,}768$-token context window. Evidence chains are read in parallel, with one call per chain and at most $6{,}000$ characters of evidence per worker. The reasoner receives each chain's draft answer and up to four supporting nodes, each truncated to $2{,}000$ characters. This bounds the evidence supplied per chain regardless of its length. Evaluation uses GPT-4o-mini at temperature $0$ with a maximum output of $300$ tokens, matching the baselines' judge settings and prompt (\hyperref[app:prompt:c1]{Prompt P7}).

The spectral allocation is concentrated: the leading direction receives a median of $73.7\%$ of the demand, with the remainder distributed across a median of $13$ additional directions (Eq.~\ref{eq:fr-aspect-capacity}). If the network's maximum flow $f_{\max}$ is below $F_0$, we reduce the routed demand to $F=\min(F_0,0.99\,f_{\max})$ to ensure feasibility. This can occur when too few content nodes can reach the aspect nodes, but does not occur on any evaluation query.

Graph construction is performed offline once per document. All controlled variants use the same graph, so their comparisons isolate changes made after graph construction.

\section{Proofs}
\label{app:proofs}

\paragraph{Proposition~\ref{prop:coverage}(i).}
The equality constraints in Eq.~\ref{eq:fr-flow} are defined by the node--arc incidence matrix of $G^+$. This matrix is totally unimodular, and adding the identity rows associated with the bounds $0\le f\le U$ preserves this property. Hence, when $b$ and $U$ are integral, every vertex of the feasible polytope is integral. For a feasible instance, the finite capacity bounds make the polytope bounded, so the linear objective attains its optimum at a vertex. An integral optimum therefore exists. Minimum-cost flow admits exact polynomial-time algorithms~\citep{ahuja1993network}. Fractional aspect capacities do not affect linearity or the conclusions of parts (ii) and (iii), although an integral optimum is no longer guaranteed.

\paragraph{Proposition~\ref{prop:coverage}(ii).}
Content arcs follow a strict total order defined by relevance, with ties broken by node index. Let $\operatorname{rank}(v)\in\{1,\ldots,|V|\}$ denote a content node's position in this order. Assign potential $0$ to $s$, $1+\operatorname{rank}(v)$ to each content node, $|V|+2$ to each aspect node, and $|V|+3$ to $t$. Every arc strictly increases this potential, so $G^+$ is acyclic.

By the flow decomposition theorem~\citep{ahuja1993network}, any feasible flow can therefore be decomposed into source-to-sink path flows without cycle flows. Each decomposition step subtracts the minimum positive flow along a path, reducing the residual flow on at least one arc to zero. Thus, at most $|E^+|$ paths are required.

\paragraph{Proposition~\ref{prop:coverage}(iii).}
Only aspect nodes connect directly to $t$, so all routed flow passes through aspect-to-sink arcs. When $F=F_0=\sum_a d_a$, flow conservation gives
\[
    \sum_a f_{at}=F_0=\sum_a d_a.
\]
Since $f_{at}\le d_a$ for every aspect, this equality requires $f_{at}=d_a$ for all $a$. Conservation at each aspect node then gives $\sum_v f_{va}=d_a$. Each content-to-aspect arc has unit capacity, so carrying $d_a$ units requires at least $\lceil d_a\rceil$ arcs with positive flow. These arcs originate from distinct content nodes, proving the claim. \hfill$\square$

\section{Path Selection by Replicator Dynamics}
\label{app:rd}

We treat each candidate chain $\pi_1,\ldots,\pi_K$ as a set of content nodes and define its quality, pairwise overlap, and selection payoff as
\begin{equation}
Q_i=\Bigl(\frac{1}{|\pi_i|}\sum_{v\in\pi_i}h_v^2\Bigr)^{1/2},
\qquad
R_{ij}=\frac{|\pi_i\cap\pi_j|}{|\pi_i\cup\pi_j|},
\qquad
M_{ij}=
\begin{cases}
Q_i^2, & i=j,\\
\sqrt{Q_iQ_j}\,(1-R_{ij}), & i\ne j.
\end{cases}
\label{eq:fr-payoff}
\end{equation}
The diagonal terms reward individual chain quality, while the off-diagonal terms favour relevant chains with few shared nodes. For node-disjoint chains, $R_{ij}=0$, so the pairwise payoff depends only on quality. The Jaccard term captures node overlap, not semantic redundancy.

Starting from a uniform distribution over candidates, $x_i^{(0)}=1/K$, replicator dynamics updates their shares as
\begin{equation}
x_i^{(t+1)}
=
x_i^{(t)}
\frac{(Mx^{(t)})_i}
     {x^{(t)\top}Mx^{(t)}}.
\label{eq:app-rd}
\end{equation}
For the symmetric, nonnegative payoff matrix $M$, the standard update monotonically increases, or leaves unchanged, the average payoff $x^\top Mx$ whenever the denominator is positive. This provides a continuous optimization approach to selecting compatible, high-quality chains~\citep{pelillo1998replicator,bomze2000approximating}.

We anneal the dynamics to sharpen selection progressively. Candidates whose shares fall below a fixed threshold relative to the uniform share $1/K$ are removed, and the remaining distribution is renormalised. Surviving chains are read in descending order of their final shares, up to the read cap. Greedy selection instead ranks chains by their routed flow. The two methods achieve similar accuracy at the reported configuration, while replicator dynamics leads by $4.1$ points when the candidate pool is doubled (Section~\ref{sec:ablate}).

\section{Evaluation Protocol}
\label{app:protocol}

\paragraph{Judge and runs.}
All reported answer accuracies are scored by GPT-4o-mini at temperature $0$ using the evaluation prompt of \citet{du2026g}, reproduced without modification in \hyperref[app:prompt:c1]{Prompt P7}. The judge agrees with its own scores on $96$--$99\%$ of identical predictions evaluated repeatedly. The three runs in Table~\ref{tab:main} were launched on different days, with serving infrastructure restarted between runs, and yield a sample standard deviation of $0.35$ macro accuracy points. Given the subset sizes, the estimated binomial standard error of macro accuracy is approximately one point.

\paragraph{Ablation samples.}
Following \citet{du2026g}, we use the same $250$ randomly sampled examples released with G\textsuperscript{2}-Reader, drawn from the five subsets. Evaluating each reported configuration on the remaining examples changes its score by at most $0.44$ points and preserves all rankings in Tables~\ref{tab:main} and~\ref{tab:ablation}. The demand cap was fixed before any hyperparameter sweep.

\paragraph{Prompts.}
We refined the prompts using the ablation examples and failure analysis. One illustrative example in the worker prompt (\hyperref[app:prompt:b1]{Prompt P4}) includes entities that also appear in three FetaTab evaluation questions. Any direct effect confined to these questions is bounded by one accuracy point on FetaTab and $0.2$ points on the macro average.

\section{Reader Cost Accounting}
\label{app:cost}

Table~\ref{tab:cost} reports the cost of reading for each router at the $16$-chain budget, measured from the runs of Table~\ref{tab:ablation}. Every variant makes the same number of model calls, $16$ parallel worker calls plus a reasoner and a query classifier, so the routers differ only in how much evidence each call carries. Wall-clock time was measured on a shared cluster and indicates an order of magnitude; tokens and calls are exact.

\begin{table}[h]
\centering
\small
\setlength{\tabcolsep}{6pt}
\begin{tabular}{@{}lrrrrr@{}}
\toprule
\textbf{Router} & \textbf{Nodes} & \textbf{Prompt tok.} & \textbf{Compl.\ tok.} & \textbf{Calls} & \textbf{s/query} \\
\midrule
Min-cost flow (\flowreader) & 17.5 & 46{,}138 & 1{,}270 & 18.0 & 25.4 \\
$k$-shortest paths          & 11.8 & 50{,}576 & 1{,}697 & 18.0 & 34.4 \\
Shortest path               & 16.9 & 63{,}106 & 2{,}002 & 18.0 & 89.9 \\
\bottomrule
\end{tabular}
\caption{\textbf{Cost of reading per query.} Prompt and completion tokens are summed over every model call of a query. Among routers that form chains, min-cost flow spends the fewest tokens and seconds per query.}
\label{tab:cost}
\end{table}

\section{Chain Structure}
\label{app:chains}

Table~\ref{tab:chains} reports the length distribution of the chains actually read. Under min-cost flow, most chains are single nodes and the remainder form a tail up to six nodes and beyond; multi-node chains nonetheless reach $88.5\%$ of queries, so structure is used where the content supports it. Shortest path, which is not bounded by aspect capacities, builds long chains on nearly every query and spends $90.3\%$ of its reads inside them, which is the source of its higher token cost in Table~\ref{tab:cost}.

\begin{table}[h]
\centering
\small
\setlength{\tabcolsep}{6pt}
\begin{tabular}{@{}lrrrr@{}}
\toprule
& & \textbf{Mean} & \textbf{Single-} & \textbf{Queries with} \\
\textbf{Router} & \textbf{Chains} & \textbf{length} & \textbf{node} & \textbf{multi-node} \\
\midrule
Min-cost flow (\flowreader) & 35{,}286 & 1.46 & 75.8\% & 88.5\% \\
$k$-shortest paths          & 35{,}099 & 1.65 & 47.5\% & 94.7\% \\
Shortest path               & 35{,}044 & 3.09 & 30.0\% & 96.0\% \\
\bottomrule
\end{tabular}
\caption{\textbf{Chain length distribution} over the full evaluation sets of \citet{du2026g}, with at most $16$ chains read per query. For min-cost flow the chain counts by length are $26{,}760$ (one node), $4{,}205$ (two), $2{,}429$ (three), $1{,}048$ (four), $490$ (five), $214$ (six) and $140$ (seven or more).}
\label{tab:chains}
\end{table}

\section{Refusal Rate across Selectors}
\label{app:refusal}

\begin{figure}[h]
    \centering
    \includegraphics[width=\textwidth]{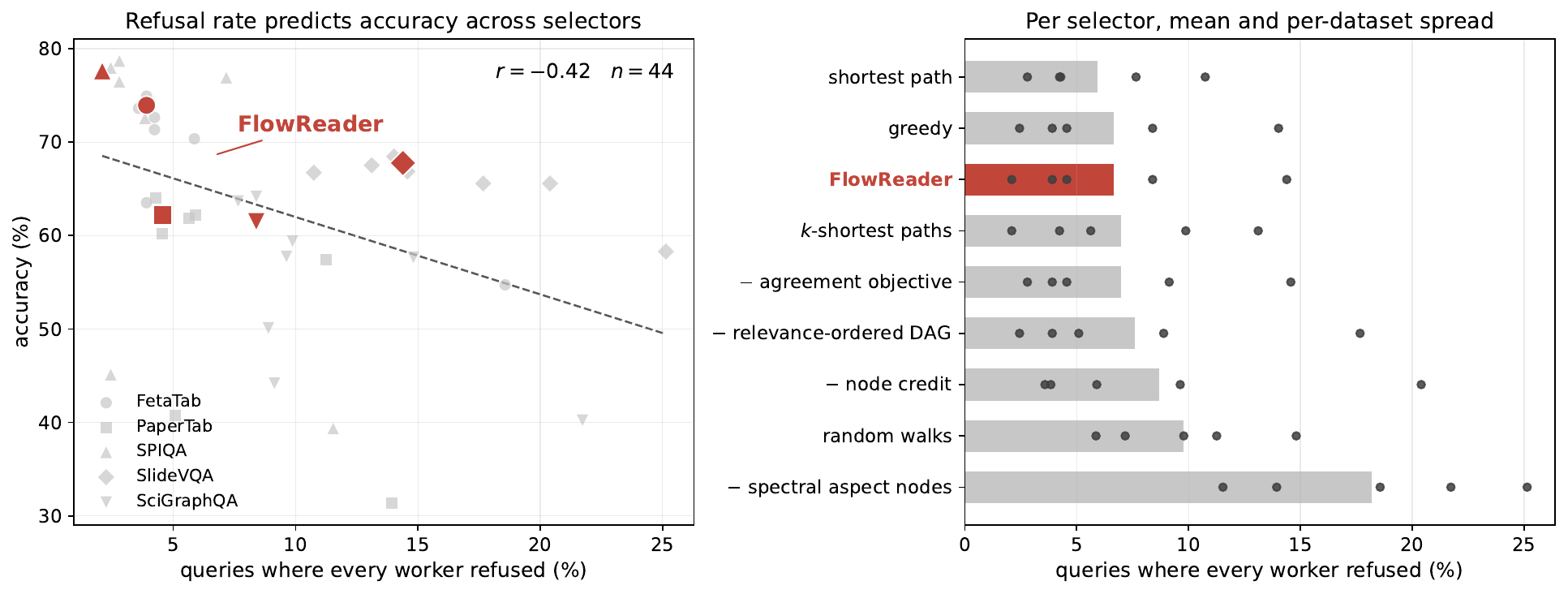}
    \caption{\textbf{Refusal rate predicts accuracy.} \emph{Left:} each point is one selector on one subset; the share of queries on which every worker refused predicts accuracy ($r=-0.42$, $n=44$). \emph{Right:} per selector, the mean and per-subset spread. Removing the spectral aspect nodes raises the all-refused rate $2.6$-fold, from $7.0\%$ to $18.2\%$, far more than any other change.}
    \label{fig:refusal}
\end{figure}

Figure~\ref{fig:refusal} relates, for each selector and subset, the share of queries on which every worker refused to accuracy. The refusal rate measures only whether the selector put answerable content in front of the readers, independently of reading and judging, and it predicts accuracy across selectors.

\section{Ablation Diagnostics}
\label{app:ablation-diag}

Each cell of Table~\ref{tab:ablation} is a single run, but every variant answers the same queries as the reported configuration, so each row is tested against it by McNemar's exact test on the paired judgements; for scale, three independent runs of the reported configuration differ by $0.35$ macro points. Table~\ref{tab:abl-diag} also reports what each variant reads. \emph{Reached} is the share of each query's sixteen most relevant nodes, by the relevance of Eq.~\ref{eq:fr-relevance}, that a variant reads, and \emph{all refused} the share of queries on which every worker returned no answer. Accuracies are pooled over queries and therefore differ slightly from the macro averages of Table~\ref{tab:ablation}.

\begin{table}[h]
\centering
\small
\setlength{\tabcolsep}{6pt}
\begin{tabular}{@{}lrrrr@{}}
\toprule
\textbf{Variant} & \textbf{$\Delta$ acc.} & \textbf{McNemar $p$} & \textbf{Reached} & \textbf{All refused} \\
\midrule
Min-cost flow (reported)          & ---    & ---               & 83.2\% & 7.0\% \\
Shortest path                     & $-0.2$ & $0.86$            & 73.9\% & 6.2\% \\
$k$-shortest paths                & $-1.0$ & $0.19$            & 63.6\% & 7.2\% \\
Replicator dynamics ($16$ reads)  & $0.0$  & $1.00$            & 83.2\% & 7.0\% \\
Greedy ($16$ reads)                   & $+0.3$ & $0.68$            & 83.6\% & 7.0\% \\
$-$ agreement objective           & $-3.1$ & $1\times10^{-4}$  & 80.8\% & 7.3\% \\
$-$ node credit                   & $-2.6$ & $0.001$           & 81.0\% & 9.3\% \\
$-$ relevance order               & $-16.1$& $<10^{-50}$       & 81.7\% & 8.1\% \\
$-$ aspect nodes (single sink)    & $-24.1$& $<10^{-50}$       & 63.3\% & 18.2\% \\
\bottomrule
\end{tabular}
\caption{\textbf{Paired significance and evidence reached} for each variant of Table~\ref{tab:ablation}, against the reported configuration. \emph{All refused} is pooled over queries. Replicator dynamics with $16$ reads fixed is judged differently from the reported configuration on only $115$ queries of the evaluation sets of \citet{du2026g}, which split $58$ to $57$, so pooled over queries the two coincide although their macro averages in Table~\ref{tab:ablation} differ by $0.2$ points.}
\label{tab:abl-diag}
\end{table}

The single-sink variant reads more distinct nodes than the reported configuration ($18.7$ against $17.5$) yet reaches markedly less of the most relevant evidence, and every worker refuses on $2.6$ times as many queries. Within the variant, accuracy follows the evidence reached: it is $19.7\%$ on queries where it reaches under a quarter of the sixteen most relevant nodes, $42.1\%$ between a quarter and a half, and $50.1\%$ between a half and three quarters. The calibrated termination scale is what keeps the flow on that evidence. The document-order variant reaches nearly as much of it as the reported configuration with a similar refusal rate, so its loss arises after reading rather than in the evidence selected.

\section{Sensitivity Analysis}
\label{app:sensitivity}

\subsection*{Retrieval demand}

Table~\ref{tab:demand} varies the relative relevance threshold $\tau$ in Eq.~\ref{eq:fr-demand}, which determines which nodes contribute to the effective count used to derive demand. All other settings remain fixed. Each row evaluates the full dataset, and node budgets are measured from the chains actually read.

Reducing the node budget by $26\%$ lowers macro accuracy by $0.9$ points. Increasing the budget from $12.9$ to $17.5$ nodes improves SciGraphQA by $3.7$ points and SlideVQA by $2.0$, while FetaTab and PaperTab change by at most $1.3$ points. These results support the reported operating point of $17.5$ nodes, the smallest tested budget that avoids a loss on any subset. Below $12.9$ nodes, accuracy declines on every subset.

\subsection*{Remaining constants}

We vary three constants individually around their reported values: the spectral energy threshold $\eta=0.90$, the reference boost $\rho_{\mathrm{ref}}=0.30$, and the termination balance $b=0.634$. All sweeps use the same $250$ ablation examples. A tenth configuration repeats the reported settings on these examples to provide a matched comparison (Table~\ref{tab:constants}). These experiments assess local sensitivity without changing the results in Table~\ref{tab:main}.

At this sample size, no tested setting is statistically distinguishable from the reported configuration. Macro accuracy varies by $1.5$ points as $\eta$ ranges from $0.70$ to $0.95$, and by $1.3$ points as $\rho_{\mathrm{ref}}$ ranges from $0$ to $0.45$. The termination balance has the largest observed effect. The reported value is the best of the four tested values, with alternatives reducing accuracy by $2.3$--$3.1$ points. This parameter is one of the seven fixed constants; Eq.~\ref{eq:app-kappa} uses it to derive $\kappa$ for each query.

The reported configuration scores $69.9$ in this separate sweep, compared with $71.0$ when the full-set runs are restricted to the same examples (Section~\ref{sec:ablate}). The difference lies within the $\pm5.7$-point uncertainty interval at this sample size. We therefore compare configurations within the same run set.

The demand and aspect-count caps are active at different rates. The demand in Eq.~\ref{eq:fr-demand} reaches $F_{\max}$ for nearly every query, whereas the aspect-count cap $A\le\operatorname{round}(F_0)$ in Eq.~\ref{eq:fr-aspect-count} is active for approximately $40\%$ of queries. This cap limits the effect of increasing $\eta$, while lowering $\eta$ can still reduce the number of retained aspects.

\begin{table}[htbp]
\centering
\small
\setlength{\tabcolsep}{7pt}
\renewcommand{\arraystretch}{1.08}
\begin{tabular}{@{}llcr@{}}
\toprule
\textbf{Constant} & \textbf{Value} & \textbf{Macro (\%)} & \textbf{$\Delta$} \\
\midrule
\multirow{4}{*}{Energy threshold $\eta$}
  & $0.70$              & 69.9 & $0.0$ \\
  & $0.80$              & 70.7 & $+0.8$ \\
  & $\mathbf{0.90}$ \tiny(reported) & \textbf{69.9} & --- \\
  & $0.95$              & 69.2 & $-0.7$ \\
\midrule
\multirow{4}{*}{Reference boost $\rho_{\mathrm{ref}}$}
  & $0$                 & 71.1 & $+1.2$ \\
  & $0.15$              & 69.8 & $-0.1$ \\
  & $\mathbf{0.30}$ \tiny(reported) & \textbf{69.9} & --- \\
  & $0.45$              & 70.8 & $+0.9$ \\
\midrule
\multirow{4}{*}{Termination balance $b$}
  & $0.45$              & 66.8 & $-3.1$ \\
  & $0.55$              & 69.5 & $-0.4$ \\
  & $\mathbf{0.634}$ \tiny(reported) & \textbf{69.9} & --- \\
  & $0.75$              & 67.6 & $-2.3$ \\
\bottomrule
\end{tabular}
\caption{\textbf{Sensitivity of the three free constants.} Each is swept one at
a time on the $250$ ablation samples of \citet{du2026g}, with the reported
configuration re-run on the same samples so every comparison is on the same samples.
Macro is the unweighted mean over the five subsets. A $95\%$ interval on any of
these estimates is $\pm 5.7$ points at this sample size, so every $\Delta$ in
the table is well inside it and no setting is separable from the reported one.}
\label{tab:constants}
\end{table}

\makeatletter
\@ifundefined{newtcblisting}{%
  \PackageError{appendix_prompts}{%
    The tcolorbox listings library is missing.^^J%
    Put ALL FOUR of these in the PREAMBLE:^^J%
    \space\space\string\usepackage{xcolor}^^J%
    \space\space\string\usepackage{listings}^^J%
    \space\space\string\usepackage[most]{tcolorbox}^^J%
    \space\space\string\usepackage{amssymb}^^J%
    The [most] option is the one that is usually missing%
  }{Without it the prompt boxes cannot be built and every prompt would be
    typeset as ordinary prose.}%
}{}
\makeatother

\definecolor{pbBuild}{HTML}{2A5E8C}
\definecolor{pbRead}{HTML}{1B7A5E}
\definecolor{pbEval}{HTML}{A85A16}

\lstdefinestyle{promptlst}{
  basicstyle=\ttfamily\scriptsize,
  breaklines=true,
  breakatwhitespace=true,
  breakindent=1.2em,
  postbreak=\mbox{\textcolor{gray}{$\hookrightarrow$}\space},
  columns=fullflexible,
  keepspaces=true,
  showstringspaces=false,
  upquote=true,
  frame=none,
  aboveskip=0pt,
  belowskip=0pt,
  literate=
    {\'{}}{{\textquotesingle}}1
    {\`{}}{{\textasciigrave}}1
    {—}{{---}}1
    {’}{{'}}1
    {“}{{``}}1
    {”}{{''}}1
    {•}{{\textbullet}}1
    {…}{{\ldots}}1
    {≡}{{$\equiv$}}1
    {✅}{{$\checkmark$}}1
    {❌}{{$\times$}}1
}

\newtcblisting{promptbox}[4]{%
  listing only,
  listing style=promptlst,
  breakable,
  enhanced,
  colback=#1!3!white,
  colframe=#1,
  colbacktitle=#1,
  coltitle=white,
  boxrule=0.6pt,
  arc=2pt,
  left=6pt, right=6pt, top=5pt, bottom=5pt,
  toptitle=2pt, bottomtitle=2pt,
  fonttitle=\bfseries\small,
  title={#2\quad #3\hfill\normalfont\ttfamily\scriptsize #4},
}

\section{Prompts}
\label{app:prompts}

Every prompt on a live call path is reproduced below, verbatim from the file
the code reads. Placeholders in \texttt{\$NAME\$} form are substituted at call
time; the source of each box is printed in its title bar so that any of them
can be checked against the repository. Prompts that only an ablation uses are
marked as such, and prompts that the repository defines but no call site
reaches are not reproduced.

\subsection*{Graph construction}

Called once per document, offline, before any query is seen.

\paragraph{P1\quad Text node analysis}\label{app:prompt:a1} Issued once per text chunk while the content graph is built. Its JSON output becomes the node's summary, keywords and tags, which are what the node embedding is computed from.

\begin{promptbox}{pbBuild}{P1}{Text node analysis}{config/prompts/text\_analysis\_keyword.txt}
Your task is to extract ALL concrete entities that explicitly appear in the provided text.

STRICT EXTRACTION RULES:
1. You MUST extract *every* entity that appears in the text.
   - Do not skip or ignore any entity, even if it seems minor.
   - Coverage must be exhaustive.

2. Only extract entities that appear *verbatim* or as clear noun phrases in the text:
   - Named entities (people, organizations, locations)
   - Technical terms and domain-specific terminology
   - Products, models, components
   - Proper nouns and unique identifiers
   - Explicit noun phrases directly present in the text

3. You MUST NOT:
   - Invent new terms
   - Generalize, abstract, or summarize
   - Produce synonyms or inferred concepts
   - Omit any entity actually present
   - Extract raw mathematical formulas, LaTeX expressions, or equations (e.g., $x_{i}$, \alpha, etc.)
   - Include special characters like $, \, {, } in keywords

4. The keywords MUST be entities present in the text, not themes.

5. Preserve the original wording of the entities from the text, but convert mathematical notation to plain text descriptions (e.g., "variable x_i" instead of "$x_{i}$").

6. When the provided content contains no meaningful information, such as reference lists, \
the summary you generate must include exactly "No meaningful information" to aid filtering.

OUTPUT FORMAT (JSON):
{
  "keywords": [
    // ALL entities that appear in the text, in their exact textual form
  ],
  "summary": "One concise sentence summarizing the content.",
  "tags": [
    // Broad classification tags, not entities
  ]
}

------------------------------------------------------------

FEW-SHOT EXAMPLES
------------------------------------------------------------

Example 1
Input Text:
"The Tesla Model S uses a Dual Motor system and features an Autopilot module.
It integrates an NVIDIA Drive PX processor and a Panasonic 18650 battery pack."

Output JSON:
{
  "keywords": [
    "Tesla Model S",
    "Dual Motor system",
    "Autopilot module",
    "NVIDIA Drive PX processor",
    "Panasonic 18650 battery pack"
  ],
  "summary": "The text describes components and hardware modules used in the Tesla Model S.",
  "tags": ["Technology", "Automotive", "Specifications"]
}

Example 2
Input Text:
"The DeepMind AlphaFold system achieved state-of-the-art results in the CASP13 competition.
The project also used the UniProt database and the PDB structure repository for training."

Output JSON:
{
  "keywords": [
    "DeepMind AlphaFold system",
    "CASP13 competition",
    "UniProt database",
    "PDB structure repository"
  ],
  "summary": "The text describes AlphaFold's performance in CASP13 and the datasets used for its training.",
  "tags": ["Machine Learning", "Biology", "Research"]
}
\end{promptbox}

\paragraph{P2\quad Visual node analysis}\label{app:prompt:a2} Issued once per figure, chart or table image, with the surrounding text and the caption substituted for \texttt{\$context\$} and \texttt{\$caption\$}. It is what makes a visual node comparable to a text node in the same embedding space.

\begin{promptbox}{pbBuild}{P2}{Visual node analysis}{config/prompts/image\_analysis\_ocrkeyword.txt}
Generate a structured analysis of the visual elements in the provided image (which is a page or figure from a scientific paper).
You are also provided with the image's surrounding text context (before and after the image) and its caption.

Instructions:
1) Visual Focus: Use the provided context only to aid your understanding of what the image conveys. The summary should be primarily based on the visual evidence and the caption.
2) Keywords: Extract exact in-image terms: labels, legends, axis titles, category names, and domain-specific terms observed in the image. Include any readable text from the image (axis numbers, annotations, data values) as keywords for search indexing.
3) Summary: When a caption is present and includes an index (e.g., "Figure 1", "Table 2"), begin the summary by formalizing it (e.g., "Figure X — …"). Describe only what is visually present and what the caption explicitly states.
4) Tags: Broad categories/themes for classification. At least three tags.
5) Text Content: For the "text_content" field, you must COMBINE the full caption text AND a transcription of all readable text found inside the image (axis numbers, labels, annotations) into a single raw string. This is for search indexing.

Format the response as a JSON object:
{
    "keywords": [
        // Include exact in-image labels, legends, axis titles, category names.
        // Also include readable text from the image (annotations, data values).
        // Order from most to least important.
        // At least three keywords.
    ],
    "summary":
        // If caption provides an index, start with "Figure X — ..." or "Table Y — ...".
        // Concisely describe the visual elements and trends.
    ,
    "tags": [
        // Broad categories/themes for classification (e.g., "Bar Chart", "Ablation Study").
        // At least three tags.
    ],
    "text_content":
        // Combine the full caption text AND all readable text from the image into one string.
        // This is used for search indexing.

}

// Few-shot Examples:

Example 1: Map with labeled points
Provided:
- Context: Describes site selection background.
- Caption: "Figure 1. Distribution of information centers and viewpoints across the region."
- Visible elements: Map showing markers labeled "Information Center" and "Viewpoint", scale bar "10km".
Expected JSON:
{
  "keywords": ["Figure 1", "Information Center", "Viewpoint", "Distribution Map", "Geospatial", "Region", "10km"],
  "summary": "Figure 1 — Distribution of information centers and viewpoints across the region. The figure presents a regional map with labeled markers indicating locations of information centers and viewpoints, highlighting spatial clustering.",
  "tags": ["Geospatial", "Map", "Urban Planning"],
  "text_content": "Figure 1. Distribution of information centers and viewpoints across the region. Information Center Viewpoint 10km"
}

Example 2: Table ablation study
Provided:
- Context: Discusses model variants.
- Caption: "Table 2. Ablation study of model components."
- Visible elements: Table columns "Method", "Accuracy", "F1-Score". Rows: "Baseline" (85.2), "Ours" (88.9).
Expected JSON:
{
  "keywords": ["Table 2", "Ablation Study", "Accuracy", "F1-Score", "Baseline", "Method", "85.2", "88.9"],
  "summary": "Table 2 — Ablation study of model components. The table compares the Baseline method against the proposed method ('Ours'), showing improvements in Accuracy and F1-Score.",
  "tags": ["Experiment", "Table", "Performance Metrics"],
  "text_content": "Table 2. Ablation study of model components. Method Accuracy F1-Score Baseline 85.2 Ours 88.9"
}

This is context: {context} and caption: {caption}
\end{promptbox}

\paragraph{P3\quad Node linking}\label{app:prompt:a3} Issued when a new node is inserted, against its nearest existing neighbours. The links it returns are the edges of the content graph the flow is later routed over. Run for three iterations throughout.

\begin{promptbox}{pbBuild}{P3}{Node linking}{config/prompts/memory\_evolve.txt}
You are an AI memory evolution agent responsible for managing and evolving a knowledge base.
Here is a memory note (including its content, summary, and keywords), and a few neighboring notes. 
Make decisions about its evolution.  

The memory note content: {content}
Summary: {context}
Keywords: {keywords}

The {neighbor_number} neighboring notes:
{neighbors}

Based on this information, determine:
1. Which of the neighboring notes should be linked to this memory note? 
2. Should the summary and keywords of this memory note be updated, considering its relationships with the neighboring notes? 
3. If so, what should be the new summary and keywords?
If you determine that the memory note needs not be updated, the new summary and keywords should be the same as the original summary and keywords. 
Please keep the summary concise, informative and to the point, with no more than 30 words.

**Definition of Valid Edge Relationships:**
Two memory notes should be connected ONLY if they exhibit one or more of the following specific logical relationships:

1. **Direct Reference/Citation**: One note explicitly references, cites, or builds upon concepts from another note.
2. **Causal Relationship**: One note describes a cause and another describes its effect, or vice versa.
3. **Part-Whole Relationship**: One note describes a component/part and another describes the whole system, or they describe different parts of the same whole.
4. **Conceptual Elaboration**: One note provides detailed explanation or elaboration of a concept introduced in another note.
5. **Temporal Sequence**: Notes describe sequential steps in a process, chronological events, or stages in a workflow.
6. **Contrastive/Comparative**: Notes discuss the same topic from different perspectives, compare alternatives, or contrast different approaches.
7. **Hierarchical Relationship**: One note describes a general concept and another describes a specific instance or subcategory.
8. **Contextual Dependency**: Understanding one note significantly enhances or requires understanding of another note.

**Exclusion Criteria - DO NOT connect notes if they only:**
- Share common keywords or terminology without substantive conceptual overlap
- Belong to the same broad domain but discuss unrelated specific topics
- Have similar surface-level similarity but no meaningful logical relationship
- Are merely adjacent in the document structure without semantic connection
- Share only generic concepts (e.g., both mention "data" or "analysis" without deeper connection)

**Connection Quality Assessment:**
When evaluating connections, ask yourself:
- Would retrieving one note significantly help understand or contextualize the other?
- Is there a specific, identifiable logical relationship beyond mere topic similarity?
- Would an expert in the domain recognize this as a meaningful connection?
- Could this connection improve retrieval accuracy for relevant queries?
- Are the suggested connections ordered according to relevancy? The most relevant connection should be the first one.

**Summary and Keywords Update Principles:**
When updating the summary and keywords, your goal is to make the memory note more precise and distinctive:
1. **Use neighbors as reference, not as content**: Compare the current note with its connected neighbors to identify what makes this note unique. However, the summary itself should ONLY describe the content of the current note. Do NOT include comparative language such as "unlike other notes", "different from neighbors", "this note does not cover X", or any explicit references to other notes.
2. **Focus on unique aspects**: If the current summary is too generic and overlaps with neighbor summaries, refine it by emphasizing the specific, unique aspects that are present in THIS note's content. 
3. **Distinctive keywords**: Update keywords to include terms that uniquely identify this note's specific content. Prefer specific, distinctive keywords over broad, generic ones that many notes might share.
4. **Self-contained description**: The summary should be a self-contained description of the current note's content. It should be clear and informative on its own, without requiring knowledge of other notes to understand it.
5. **When NOT to update**: If the current summary already accurately captures the unique aspects of this note and clearly distinguishes it from neighbors (through its specificity, not through comparative language), do not update it.

**Summary Writing Guidelines:**
- ✅ DO: "This note discusses the implementation details of algorithm X, focusing on its time complexity analysis."
- ✅ DO: "This note describes the experimental setup for testing hypothesis Y, including specific parameters and measurement protocols."
- ❌ DON'T: "Unlike other notes, this note focuses on algorithm X rather than algorithm Y."
- ❌ DON'T: "This note does not cover theoretical foundations, which are discussed in other notes."
- ❌ DON'T: "This note is different from its neighbors in that it analyzes..."

Return your decision in JSON format with the following structure:
{
    "suggested_connections": ["neighbor_memory_ids"], 
    "should_update": True or False,
    "new_summary": "new summary",
    "new_keywords": ["keyword_1",..."keyword_n"]
}
\end{promptbox}

\subsection*{Evidence reading and answer synthesis}

Called per query, after the flow has been routed and the chains selected.

\paragraph{P4\quad Chain reader (worker)}\label{app:prompt:b1} Issued once per selected evidence chain, in parallel. \texttt{\$DOC\$} receives that chain's nodes in routed order and \texttt{\$Q\$} the query. This is the only prompt that sees document content at inference time.

\begin{promptbox}{pbRead}{P4}{Chain reader (worker)}{config/prompts/worker\_noabstain.txt}
Read the evidence passages and images below, then answer the question.

<text>
$DOC$
</text>

Question: $Q$

ANSWERING POLICY (read this first):
Your job is to report what this evidence says about the question, even when it
is partial. Another stage decides whether your answer is well supported and
compares it against other readers' answers, so a partial or uncertain answer is
useful and a refusal is not. Refusing costs the system the answer entirely.

- If the evidence gives the answer, give it.
- If the evidence gives PART of the answer — one of two values asked for, a
  related row, a number without its label, a trend without its magnitude —
  report that part. Do not refuse because it is incomplete.
- If the evidence only lets you narrow it down, give the best-supported
  candidate and quote what you based it on.
- Use "not_reported" ONLY when the passage is about a clearly different subject
  than the question, or shares no relevant content at all. Uncertainty is not a
  reason to refuse; irrelevance is.

Rules:
- Use only the provided text and images. Never answer from general knowledge.
- ENTITY CHECK: identify the primary subject(s) of the question — proper nouns
  (people, films, companies, datasets, models, papers, places, products). If the
  evidence describes a DIFFERENT entity (the question asks about
  "Aaron Taylor-Johnson" and the passage describes T.J. Miller's roles), that is
  the one case for "not_reported": do not substitute a plausible-looking fact
  from the wrong entity. But evidence that mentions the right subject only
  briefly, or in a caption or a single table row, IS usable — report what it
  says.
- For tables: when the question asks for a specific value, match the exact
  row/column intersection.
- For tables: when the question asks for "improvements", "gains", "how much
  better", or "increase over baseline", read ALL relevant rows (target model AND
  baseline rows) and compute the difference. Report delta values (e.g. "+0.05"),
  not absolute values.
- For "which experiment/method/model": report the label, not the numbers.
- For "which season/episode/edition/year/version performed best / is highest /
  is largest": report the LABEL (e.g. "season two", "2015 edition"), NOT the
  numeric value from its row.
- For questions that ask which of two named options is higher/better/superior
  ("in A or B?", "X or Y?", "single-domain or multi-domain?", "which setting
  achieves higher"): extract the raw metric value for EACH named option, compare
  them directly, and name the winning option. Do NOT decide based on which
  option achieves new state-of-the-art relative to prior work — compare the
  absolute values of the two options only.
- DIRECTION / POLARITY (critical): for questions about trends, changes, or
  relationships (increases/decreases, improves/worsens, higher/lower,
  positive/negative, rises/falls, correlates/inversely): (1) locate the exact
  values or explicit direction words in the evidence, (2) state the direction you
  read — do NOT infer from expectation or general knowledge, (3) then give the
  answer. If the evidence shows X goes from 0.6 to 0.8, the direction is
  "increases" regardless of what is typical.
- BOOLEAN (critical): for yes/no questions, restate the exact relevant fact from
  the evidence before deciding. Never answer Yes/No from general knowledge — only
  from the evidence in this context. If the evidence bears on the question but
  does not settle it, answer with the direction it points and quote the fact.

Respond with ONLY a JSON object, no other text:
{"status":"supported","answer":"<exact answer>","evidence":"<key phrase from text>"}
or, only for a clearly different subject or no relevant content:
{"status":"not_reported","answer":"","evidence":""}
\end{promptbox}

\paragraph{P5\quad Reconciliation (reasoner)}\label{app:prompt:b2} Issued once per query, after the workers return. \texttt{\$PATHS\$} receives the per-chain drafts and the evidence behind them. It produces the answer that is scored.

\begin{promptbox}{pbRead}{P5}{Reconciliation (reasoner)}{config/prompts/reasoner\_shape\_trust.txt}
Answer the question from the evidence paths below.

Each path gives a **worker answer** and its **key evidence**. The worker read
the FULL passage for that path, including any images, tables and figures. The
key evidence below is a short text excerpt of that passage — it is often
incomplete, and for slides, charts and tables it frequently does not contain
the number or label the worker read off the image.

QUESTION
$Q$

EVIDENCE PATHS
$PATHS$

How to weigh workers against excerpts:
1. A worker answer is wrong only if some key evidence CONTRADICTS it — shows a
   different value, entity or direction for the same thing. An excerpt that is
   merely silent about the fact does not contradict the worker: the worker saw
   more than the excerpt shows. Never replace a worker's answer with "not
   reported" because you cannot see the fact in the excerpt yourself.
2. When several workers independently give the same answer, that agreement is
   strong evidence. Use it unless an excerpt directly contradicts it.
3. Where a worker refused but a path's key evidence contains the requested
   fact, extract it yourself.
4. Check the question's subject. Ignore paths whose key evidence is about a
   different entity, however plausible their answer looks.
5. Combine complementary partial findings across paths into one answer. If
   the question asks for several things, answer each; do not drop a part a
   worker supplied because another part is missing.
6. Only say the question cannot be answered if NO worker answered and NO
   excerpt contains the fact.

ANSWER FORM. Write the answer as a self-contained statement that a reader who
has not seen the question would understand: name the subject the question asks
about and state the fact. Never answer with a bare label, number, name, or
Yes/No on its own — when a worker's answer is bare, complete it from the
evidence.
   - yes/no question: start with exactly "Yes" or "No", then in the same
     sentence the fact that decides it.
   - names two options and asks which is higher/better: name the winner and
     give both values if the evidence has them, not Yes/No.
   - asks for a value: the subject, the value and its unit, exactly as written
     in the source.
   - asks how much something improved or differs: the signed difference, and
     the two absolute values it comes from.
   - asks which method/entity/item: the label, plus the property that makes it
     the answer (e.g. "X, with the lowest error of 3.1%").
   - asks to list: every item the evidence supports, none dropped. If the
     question mentions a comparison or a reference, give both sides.
   - asks what a figure or table shows, what trend, relationship, correlation
     or observation it presents, or how one thing compares to another: state
     the direction of the relationship (what increases, decreases, or stays
     the same, and relative to what), name the quantities and the conditions,
     and then the reason if the evidence gives one. Two or three sentences.
     Never only describe what is plotted or which axes it has.
   - asks how, why, describe or explain: a short paragraph that covers each
     component the question names, grounded in the key evidence.
   Copy the source's exact wording for values, labels and names. Do not add
   facts the question did not ask about, and do not pad with generalities.

Wrap your reasoning in <thought>...</thought>, then your final answer in
<output>...</output>. The <output> tag must be the last thing you write, and
must contain no markdown, image references or nested tags.

<thought>
[2-5 sentences]
</thought>
<output>your answer</output>
\end{promptbox}

\paragraph{P6\quad Sub-question aspect source}\label{app:prompt:b4} Not part of the reported system. Used by the aspect-source ablation of Section~\ref{sec:ablate}, which replaces the spectral directions with generated sub-questions while holding the rest of the pipeline fixed.

\begin{promptbox}{pbRead}{P6}{Sub-question aspect source}{flowreader/prompts.py :: subquestion\_prompt}
Break the question below into its distinct parts -- the separate things that
must each be found in the document for the question to be fully answered.

Rules:
- At most $N$ parts. Fewer is better; give one part if the question asks one
  thing.
- Each part must be answerable from the document on its own.
- Do not invent related questions. Only decompose what is actually asked.
- Keep each part short and concrete, in the question's own vocabulary.

Return JSON only: {"parts": ["...", "..."]}

Question: $Q$
\end{promptbox}

\subsection*{Evaluation}

Called once per query on the finished answer. No document content is shown to the judge.

\paragraph{P7\quad Judge}\label{app:prompt:c1} Issued once per query over (question, gold answer, predicted answer), and used for every number reported in this paper. It is the evaluation prompt of \citet{du2026g}, byte-identical to \texttt{config/prompts/eval.txt} in that work's repository, so our results and the baselines transcribed from it are scored by the same rubric as well as the same model and decoding settings (GPT-4o-mini, temperature~$0$, $300$ max tokens).

\begin{promptbox}{pbEval}{P7}{Judge}{config/prompts/eval\_g2reader.txt}
You are an expert evaluator tasked with assessing the accuracy of answers generated by a RAG
(Retrieval-Augmented Generation) system.

**Task**: Evaluate whether the generated answer correctly responds to the given question based on the expectedanswer.

**Question**: {question}

**Expected Answer**: {gold_answers}

**Generated Answer**: {assistant_answer}

**Evaluation Criteria**:
1. **Accuracy (0 or 1)**: Does the generated answer match the factual content of the expected answer?
- 1: The generated answer is factually correct and aligns with the expected answer
- 0: The generated answer is factually incorrect or contradicts the expected answer

**Instructions**:
- Focus on factual correctness, not writing style or format
- Consider partial matches: if the generated answer contains the correct information but includes additional context, it should still be considered accurate
- For numerical answers, check if the values match or are equivalent
- For list answers, check if all key elements are present
**"Not answerable" handling**: 
  - If BOTH the expected answer AND generated answer indicate "Not answerable" or inability to answer, consider it accurate (accuracy=1)
  - If the expected answer provides specific content but the generated answer says "Not answerable", it is INCORRECT (accuracy=0)
  - If the expected answer says "Not answerable" but the generated answer provides specific content, it is INCORRECT (accuracy=0)

**Output Format**:
Please respond with a JSON object containing only:
{{
"accuracy": 0 or 1,
"reasoning": "Brief explanation of your evaluation"
}}
- Output MUST be valid JSON (double quotes only).
- Do NOT use LaTeX, math delimiters, or any backslashes "\" in reasoning.
- If you need to mention N_C, write it as "N_C" or "NC" (no "\" characters).
- No code fences, no extra text.
\end{promptbox}

\end{document}